\documentclass[12pt]{article}

\usepackage{amssymb}

\newcommand{\sca}{\ensuremath{\mathcal{A}}}

\newcommand{\scd}{\ensuremath{\mathcal{D}}}
\newcommand{\scg}{\ensuremath{\mathcal{G}}}
\newcommand{\sch}{\ensuremath{\mathcal{H}}}
\newcommand{\scs}{\ensuremath{\mathcal{S}}}
\newcommand{\scx}{\ensuremath{\mathcal{X}}}

\newcommand{\sone}{\ensuremath{\mathcal{S}_{1}}}

\newcommand{\sdera}{\ensuremath{SDer(\mathcal{A})}}

\newcommand{\phst}{\ensuremath{\Phi_*}}
\newcommand{\phstup}{\ensuremath{\Phi^*}}

\newcommand{\oa}{\ensuremath{\mathcal{O}(\mathcal{A})}}

\begin{document}
\noindent \textbf{\large A Stepwise Planned Approach to the Solution
of Hilbert's Sixth Problem. II : Supmech and Quantum Systems}

\vspace{.15in}\noindent \textbf{Tulsi Dass}

\noindent Indian Statistical Institute, Delhi Centre, 7, SJS
Sansanwal Marg, New Delhi, 110016, India.

\noindent  E-mail: tulsi@isid.ac.in; tulsi@iitk.ac.in

\vspace{.18in} \noindent \textbf{Abstract:} Supmech, which is
noncommutative Hamiltonian mechanics \linebreak (NHM) (developed in
paper I) with two extra ingredients : positive observable valued
measures (PObVMs) [which serve to connect state-induced expectation
values and classical probabilities] and the `CC condition' [which
stipulates that the sets of observables and pure states be mutually
separating] is proposed as a universal mechanics potentially covering
all physical phenomena. It facilitates development of an autonomous
formalism for quantum mechanics. Quantum systems, defined algebraically
as supmech Hamiltonian systems with non-supercommutative system
algebras, are shown to inevitably have Hilbert space based
realizations (so as to accommodate rigged Hilbert space based Dirac
bra-ket formalism), generally admitting commutative superselection
rules. Traditional features of quantum mechanics of finite particle
systems appear naturally. A treatment of localizability much simpler
and more general than the traditional one is given. Treating
massive particles as localizable elementary quantum systems, the
Schr$\ddot{o}$dinger wave functions with traditional Born
interpretation appear as natural objects for the description of
their pure states and the Schr$\ddot{o}$dinger equation for them
is obtained without ever using a classical Hamiltonian or Lagrangian.
A provisional set of axioms for the supmech program is given.

\newpage
\noindent \textbf{I. Introduction}

\vspace{.1in} This is the second of a series of papers aimed at
obtaining a solution of Hilbert's sixth problem in the framework  of
a noncommutative geometry (NCG) based `all-embracing' scheme of
mechanics. In the first paper (Dass [15]; henceforth referred to as
I), the `bare skeleton' of that mechanics was presented in the form
of noncommutative Hamiltonian mechanics (NHM) which combines
elements of noncommutative symplectic geometry and noncommutative
probability in the setting of topological superalgebras. Consideration
of interaction between two systems in the NHM framework led to the
division of physical systems into two `worlds'
--- the `commutative world' and the `noncommutative world'
[corresponding, respectively, to systems with (super-)commutative
and non-(super-)commutative system algebras] --- with no consistent
description of interaction allowed between two systems belonging to
different `worlds'; in the `noncommutative world', the system
algebras  are constrained by the formalism to have a `quantum
symplectic structure' characterized by a universal Planck type
constant.

 The formalism of NHM presented in  I is deficient
in that it does not connect smoothly to classical probability and,
in the noncommutative case, to Hilbert space. A refined version of
it, called Supmech, is presented in section 2 which has two extra
ingredients aimed at overcoming these deficiencies.

The first ingredient is the introduction of classical probabilities
as expectation values of `supmech events' constituting `positive
observable-valued measures' (PObVMs) [a generalization of positive
operator-valued measures]. All probabilities in the formalism
relating to the statistics of outcomes in experiments are
stipulated to be of this type.

The second ingredient is the condition of `compatible completeness'
between observables and pure states (referred to as the `CC
condition') -- the condition that the two sets be mutually
separating. This condition is satisfied in classical Hamiltonian
mechanics  and in traditional Hilbert space quantum mechanics (QM).
(It is, however, not generally satisfied in superclassical
Hamiltonian systems with a finite number of fermionic generators;
see section 2.3). It will be seen to play an important role in the
whole development; in particular, it serves to smoothly connect ---
without making any extra assumptions --- the algebraically defined
quantum systems with the Hilbert space-based ones.

A general treatment of localizable systems (more general and simpler
than that in the traditional approaches), which makes use of PObVMs,
is given in section 2.4. In section 2.5, elementary systems are
defined in supmech and the special case of nonrelativistic
elementary systems is treated. The role of relativity groups in the
identification of fundamental observables of elementary systems is
emphasized. Particles are proposed to be treated as localizable
elementary systems.

In section 3, quantum systems are treated as supmech Hamiltonian
systems with non-(super-)commutative system algebras. As mentioned
above, the CC condition ensures the existence of their Hilbert space
based realizations. In the case of systems with finitely generated
system algebras, one has an irreducible faithful representation
(unique up to unitary equivalence) of
the system algebra; in the general case, one has a direct sum of
such representations corresponding to situations with commutative
superselection rules. Treating material particles as localizable
elementary quantum systems, the Schr$\ddot{o}$dinger wave functions
are shown to appear naturally in the description of pure states;
their traditional Born interpretation is obvious and the
Schr$\ddot{o}$dinger equation appears as a matter of course ---
without ever using the classical Hamiltonian or Lagrangian in the
process of obtaining it. The Planck constant is introduced at the
place dictated by the formalism (i.e. in the quantum symplectic
form); its appearance everywhere else ---
 canonical commutation relations, Heisenberg and
Schr$\ddot{o}$dinger equations, etc. --- is automatic.

In section 4, a transparent treatment of quantum - classical
correspondence in the supmech framework is presented showing
the emergence, in the $\hbar \rightarrow 0$ limit, of classical
Hamiltonian systems from the quantum systems treated as
noncommutative supmech hamiltonian systems. In
section 5, a provisional set of axioms underlying the treatment of
systems in the supmech framework is given. The last section contains
some concluding remarks.

\vspace{.15in} \noindent \textbf{2. Augmented Noncommutative
Hamiltonian Mechanics : Supmech}

\vspace{.1in} The two new ingredients for NHM  mentioned above (the
PObVMs and the CC condition) are introduced in sections 2.1 and 2.2;
section 2.3 contains an example of an NHM system violating the CC
condition. The PObVMs will be used in section 2.4 in the treatment
of localizable systems. The CC condition will be used in section 2.5
to allow the Hamiltonian action of a relativity group on the system
algebra of an elementary system to be extended to a Poisson action
(of the corresponding projective group)
which is an important simplification. Noncommutative Noether
invariants of the projective Galilean group for a free massive spinless
particle will be obtained in section 2.6.

We shall freely use the terminology and notation of I. We quickly
recall here that, in NHM, a physical system is assumed to have
associated with it a (topological) superalgebra \sca \ (with unit
element I), the even hermitian elements of which are identified as
the system observables. Observables of the form of finite sums $
\sum A_i^* A_i  \ (A_i \in \sca) $ are called positive. A state
$\phi$ of \sca \ is  defined as a (continuous) positive linear
functional on \sca \ satisfying the normalization condition $\phi
(I) = 1;$ the quantity $\phi(A)$ is to be interpreted as the
expectation value of the observable A when the system is in the
state $\phi$. Sets of observables, states and pure states (those not
expressible as nontrivial convex combinations of other states) of
\sca \ are denoted as $\oa, \scs(\sca)$ and \sone(\sca)
respectively.

\vspace{.1in} \noindent \emph{Note.} In a couple of earlier versions
of I (arXiv : 0909.4606 v1, v2), the following convention about the
*-operation in a superalgebra \sca \ [following (Dubois-Violette
[21], section 2)] was adopted :
\[ (A B)^* = (-1)^{\epsilon_A \epsilon_B} B^* A^* \]
where $\epsilon_A $ is the parity of $A \in \sca$. This convention,
however, does not suit the needs of the work reported in this series
(it was not used anywhere in I). We shall henceforth use the
convention $ (AB)^* = B^* A^* $. [Given two fermionic annihilation
operators a, b, for example, we have $(ab)^* = b^* a^*$ and not
$(ab)^* = - b^* a^*$. One can also check the appropriateness of the
latter convention by taking \sca \ to be the superalgebra of linear
operators on a superspace $ V = V^{(0)} \oplus V^{(1)}$.]

\vspace{.12in} \noindent \textbf{2.1. Positive observable valued
measures}

\vspace{.1in} We shall introduce classical probabilities in the
formalism through a straightforward formalization of a measurement
situation. To this end, we consider a measurable space $(\Omega,
\mathcal{F})$ and associate, with  every measurable set $E \in
\mathcal{F}$, a positive observable $\nu(E)$ such that
\begin{eqnarray}\begin{array}{l}
(i) \ \nu(\emptyset)=0, \ (ii) \  \nu(\Omega) = I,   \nonumber \\
(iii)\ \nu(\cup_i E_i) = \sum_i \nu(E_i)\  \textnormal{(for disjoint
unions)}.\end{array} \end{eqnarray} [The last equation means that,
in the relevant topological algebra, the possibly infinite sum on
the right hand side is well defined and equals the left hand side.]
Then, given a state $\phi$, we have a probability measure $p_{\phi}$
on $(\Omega, \mathcal{F})$ given by
\begin{eqnarray}
p_{\phi}(E) = \phi(\nu(E)) \ \ \forall E \in \mathcal{F}.
\end{eqnarray}
The family $ \{ \nu (E), E \in \mathcal{F} \}$ will be called a
\emph{positive observable-valued measure} (PObVM) on $(\Omega,
\mathcal{F})$. It is the abstract counterpart of the `positive
operator-valued measure' (POVM) employed in Hilbert space QM (Davies
[17]; Holevo [26]; Busch, Grabowski, Lahti [12]). The objects $\nu
(E)$ will be called \emph{supmech events} (representing possible
outcomes in a measurement situation); these are algebraic
generalizations of the objects (projection operators) called
`quantum events' (Parthasarathy [38]). A state assigns probabilities
to these events. Eq.(1) represents the desired relationship between
the supmech expectation values and classical probabilities.

It is stipulated that all probabilities in the formalism relating to
statistics of outcomes in experiments must be of the form (1).

In concrete applications, the space $\Omega$ represents the `value
space' (spectral space) of one or more observable quantities. The
measurable subsets of $\Omega$ (elements of $\mathcal{F}$) represent
idealised domains supposed to be experimentally accesible. In a
classical probability space $(\Omega, \mathcal{F}, P^{cl})$, they
are the `events' to which probabilities are assigned by the
probability measure $P^{cl}$; the classical probability of an event
$E \in \mathcal{F}$ is \begin{eqnarray} P^{cl}(E) = \int_{\Omega}
\chi_E dP^{cl} \equiv \phi_{P^{cl}} (\chi_E) \end{eqnarray} where
$\chi_E$ is the characteristic/indicator function of the subset E
(the random variable which represents the classical observable
distinguishing between the occurrence and non-occurrence of the
event E. [These random variables are easily seen to constitute a
PObVM on the commutative unital *-algebra $\tilde{\sca}_{cl}$ of
complex measurable functions on $(\Omega, \mathcal{F});$ the objects
$\nu (E) $ described above are noncommutative generalizations of
these.] The right hand side of (2) expresses the classical
probability of occurrence of the event E as expectation value of the
observable $\chi_E$ in the state $\phi_{P^{cl}}$ [represented by the
probability measure $P^{cl}$ on the measurable space $(\Omega,
\mathcal{F})$] of the commutative algebra $\tilde{\sca}_{cl}.$

We have here a more sophisticated scheme of probability theory which
incorporates classical probability theory as a special case and is
well equipped to take into consideration the influence of one
measurement on probabilities of outcomes of other measurements.
Moreover, this scheme appears embedded in an `all-embracing' scheme
of mechanics --- in the true spirit of Hilbert's sixth problem.

Concrete examples of the objects $\nu (E)$ will appear in sections
2.4 and 3.4 where observables related to localization are treated.

\vspace{.12in} \noindent \textbf{2.2. The condition of compatible
completeness on observables and pure states}

\vspace{.1in} In a sensible physical theory, the collection of pure
states of a system must be rich enough to distinguish between two different
observables. (Mixed states represent averaging over ignorances over
and above those implied by the irreducible probabilistic aspect of
the theory; they, therefore, are not the proper objects for a
statement of the above sort.) Similarly, there should be enough
observables to distinguish between different pure states. These
requirements are taken care of in supmech  by stipulating that the
pair (\oa, \sone(\sca)) be \emph{compatibly complete} in the sense
that

\vspace{.1in} \noindent (i) given $A,B \in \oa, A \neq B, $ there
should be a state $\phi \in \sone(\sca)$ such that $ \phi(A) \neq
\phi(B)$;

\vspace{.1in} \noindent (ii) given two different states $\phi_1$ and
$\phi_2$ in \sone(\sca), there should be an $A \in \oa$ \ such that
$ \phi_1(A) \neq \phi_2(A)$.

\vspace{.1in} \noindent We shall refer to this condition as the `CC
condition' for the pair $ (\oa, \sone(\sca))$.

\vspace{.1in} \noindent  \textbf{Proposition 2.1} \emph{The CC
condition holds for (i) a classical Hamiltonian system
$(M,\omega_{cl}, H_{cl})$ [where $(M,\omega_{cl})$ is a finite
dimensional symplectic manifold and the Hamiltonian $H_{cl}$ is a
smooth real valued function on M] and (ii) a traditional quantum
system represented by a quantum triple $ (\sch, \mathcal{D}, \sca)$
where \sch \ is a complex separable Hilbert space, $\mathcal{D}$ a
dense linear subset of \sch \ and \sca \ is an Op*-algebra based on
the pair $ (\sch, \mathcal{D})$ acting irreducibly [i.e. such that
there does not exist a smaller quantum triple $ (\sch^{\prime},
\mathcal{D}^{\prime}, \sca)$ with $\mathcal{D}^{\prime} \subset
\mathcal{D}, \sca \mathcal{D}^{\prime} \subset \mathcal{D}^{\prime}$
and $ \sch^{\prime}$ is a proper subspace of \sch].}

\vspace{.1in} \noindent [Note. Op$^*$- algebras (Horuzhy [28]) and
quantum triples were defined in section 3.4 of I.]

\vspace{.1in} \noindent  \emph{Proof.} (i) For a classical
hamiltonian system $(M,\omega_{cl}, H_{cl})$, observables are smooth
real valued functions on M and pure states are Dirac measures (or,
equivalently, points of M) $ \mu_{\xi_0}  (\xi_0 \in M);$ the
expectation value of the observable f in the pure state
$\phi_{\xi_0}$ corresponding to the Dirac measure $\mu_{\xi_0}$ is
given by $ \phi_{\xi_0}(f) = \int f d\mu_{\xi_0} = f (\xi_0)$. Given
two different real-valued smooth functions on M, there is a point of
M at which they take different values; conversely, given two
different points $\xi_1$ and $\xi_2$ of M, there is a real-valued
smooth function on M which takes different values at those points.
[To show the existence of such a function, let U be an open
neighborhood of $\xi_1$ not containing $\xi_2$; now appeal to lemma
(2) on page 92 of (Matsushima [35])  which guarantees the existence
of a smooth function non-vanishing at $\xi_1$ and vanishing outside
U.]

\noindent (ii) The observables are the Hermitian elements of \sca \
and pure states are unit rays represented by normalized elements of
$\mathcal{D}$.

\noindent(a) Given $A,B \in \mathcal{O}(\sca)$, and $ (\psi, A \psi)
= (\psi, B \psi)$ for all normalized $\psi$ in $\mathcal{D}$ (hence
for all $\psi$ in $\mathcal{D}$), we have $(\chi, A \psi) = (\chi, B
\psi)$ for all $\chi, \psi \in \mathcal{D}$, implying A = B. [Hint :
Consider the given equality with the state vectors $(\chi +
\psi)/\sqrt 2$ and $(\chi +i \psi)/\sqrt 2$.]

\noindent (b) Given  normalized vectors $ \psi_1, \psi_2 $ in
$\mathcal{D}$ and $ (\psi_1, A \psi_1) = (\psi_2, A \psi_2)$ for all
$ A \in \mathcal{O}(\sca)$, we must prove that  $ \psi_1 = \psi_2$
up to a multiplicative phase factor. Considering the 2-dimensional
subspace V of \sch \ spanned by $\psi_1$ and $\psi_2$ and choosing
an appropriate orthonormal basis in V, we can write
\[ \psi_1 = \left( \begin{array}{c}1 \\0 \end{array} \right), \ \
\psi_2 = \left( \begin{array}{c} a \\b \end{array} \right) \ \
\textnormal{with} \ |a|^2 + |b|^2 = 1. \] It is easily seen that
$\psi_2 = U \psi_1$ where (writing $a = |a| e^{i \alpha}, \ b = |b|
e^{i \beta}$) U is the unitary matrix
\[ U = \left( \begin{array}{cc} a & b e^{i(\alpha - \beta)} \\
b &  - a e^{i(\beta - \alpha)} \end{array}\right). \] Extending U
trivially to a unitary operator on \sch \ (and denoting the extended
operator by U) we again have $\psi_2 = U \psi_1$ (in \sch). The
given equality and denseness of $\mathcal{D}$ then give $U^*AU = A$
(for all $A \in \oa$, hence all $A \in \sca$). The irreducibility of
\sca-action now implies U = I up to a multiplicative phase factor. \
\ $\Box$

\vspace{.1in} \noindent \emph{Note}. The irreducibility of
\sca-action assumed above implies that all elements of $\mathcal{D}$
represent pure states. This excludes the situations when \sch \ is a
direct sum of more than one coherent subspaces in the presence of
superselection rules.

The noncommutative Hamiltonian mechanics (NHM) described in I
augmented by the two inclusions --- PObVMs and the CC condition ---
is being hereby projected as the `all-embracing' mechanics covering
(in the sense of providing a common framework for the description
of) all motion in nature; we shall henceforth refer to it as
\emph{Supmech}.

 We have seen in section 3.4 of I  that both --- classical
Hamiltonian mechanics and traditional Hilbert space quantum
mechanics ---  are subdisciplines of NHM. Since the two new
ingredients --- PObVMs and the CC condition --- are present in both
of them, both  are subdisciplines of supmech as well.

\vspace{.12in} \noindent \textbf{2.3. Superclassical systems;
Violation of the CC condition}

\vspace{.12in} Superclassical mechanics is an extension of classical
mechanics which employs, besides the traditional phase space
variables, Grassmann variables $\theta^{\alpha}\  (\alpha = 1,..n$,
say) satisfying the relations
$\theta^{\alpha}\theta^{\beta} + \theta^{\beta}\theta^{\alpha} = 0$
for all $\alpha, \beta$; in particular, $(\theta^{\alpha})^2 = 0$
for all $ \alpha$. These
objects generate the so -called Grassmann algebra (with n
generators) $ \mathcal{G}_n$ whose elements are functions of the
form
\begin{eqnarray*}
f(\theta) = a_0 + a_{\alpha} \theta^{\alpha} + a_{\alpha \beta}
\theta^{\beta} \theta^{\alpha} + ...
\end{eqnarray*}
where the coefficients $ a_{..}$ are complex numbers; the right hand
side is obviously a finite sum. If the coefficients $a_{..}$ are
taken to be smooth functions on, say, $\mathbb{R}^m$, the resulting
functions $ f(x,\theta) $ are referred to as smooth functions on the
superspace $ \mathbb{R}^{m \mid n}; $ the algebra of these functions
is denoted as $ C^{\infty}(\mathbb{R}^{m \mid n}). $ With parity
zero assigned to the variables $x^a$ (a = 1,..,m) and one to the
$\theta^{\alpha}, C^{\infty}(\mathbb{R}^{m \mid n})$ is a
supercommutative superalgebra [with multiplication given by
$ (fg) (x, \theta) = f(x, \theta) g(x, \theta)$]. Restricting the
variables $x^a$ to an
open subset U of $R^m$, one obtains the superdomain $U^{m \mid n}$
and the superalgebra $C^{\infty}(U^{m \mid n})$ in the
above-mentioned sense. Gluing such superdomains appropriately, one
obtains the objects called supermanifolds. These are the objects
serving as phase spaces in superclassical mechanics. We shall, for
simplicity, restrict ourselves to the simplest supermanifolds
$\mathbb{R}^{m \mid n}$ and take, as system algebra, $ \sca =
C^{\infty}(\mathbb{R}^{m \mid n}).$ A *-operation is assumed to be
defined on \sca \ with respect to which the `coordinate variables'
$x^a$ and $\theta^{\alpha}$ are assumed to be hermitian.

States in superclassical mechanics are normalized positive linear
functionals on $ \sca = C^{\infty}(\mathbb{R}^{m \mid n})$; they are
generalizations of the states in classical statistical mechanics
given by
\begin{eqnarray*}
\phi(f) = \int_{\mathbb{R}^{m \mid n}} f(x, \theta) d \mu (x,
\theta)
\end{eqnarray*}
where the measure $\mu$ satisfies the normalization and positivity
conditions
\begin{eqnarray}
1 = \phi (1) = \int d \mu (x, \theta); \\
0 \leq \int f f^* d \mu \hspace{.2in} \textnormal{for all} \ f \in
\sca.
\end{eqnarray}
For states admitting  a density function, we have
\begin{eqnarray*}
d \mu (x, \theta) = \rho (x, \theta) d\theta^1...d\theta^n d^mx.
\end{eqnarray*}
To ensure real expectation values for observables, $ \rho(.,.)$ must
be even (odd) for n even (odd). The condition (3) implies that
\begin{eqnarray}
\rho(x, \theta) = \rho_0(x) \theta^n...\theta^1 + \textnormal{terms
of lower order in} \ \theta
\end{eqnarray}
where $ \rho_0$ is a probability density on $ \mathbb{R}^m$.

The CC condition  is generally not satisfied by the pair $ (\oa,
\sone(\sca))$ in super-classical mechanics. To show this, it is
adequate to give an example (Berezin [8]). Taking $ \sca =
C^{\infty}(\mathbb{R}^{0 \mid 3}) \equiv \mathcal{G}_3$, we have a
general state represented by a density function of the form
\begin{eqnarray*}
\rho (\theta) = \theta^3 \theta^2 \theta^1 +
c_{\alpha}\theta^{\alpha}.
\end{eqnarray*}
The inequality (4) with $ f = a \theta^1 + b \theta^2 $ (with a and
b arbitrary complex numbers) implies $c_3 = 0;$ similarly, $ c_1 =
c_2 = 0, $ giving, finally $ \rho(\theta) = \theta^3 \theta^2
\theta^1. $ There is only one possible state which must be pure.
This state does not distinguish, for example, observables $ f = a +
b \theta^1 \theta^2 $ with the same `a' but different `b', thus
verifying the assertion made above.

\vspace{.1in} \noindent \emph{Note.} It would not do to stipulate
exclusion of $\theta$-dependence in observables. Treatments in
superclassical mechanics,  of particles with spin, for example,
employ $\theta$-dependent observables (Berezin [8], Dass [14]).

\vspace{.1in} Superclassical  mechanics with a finite number of odd
variables, therefore, appears to have a fundamental inadequacy; no
wonder, therefore, that it does not appear to be realized by systems
in nature. The argument presented above, however, does not apply to
the $ n = \infty$ case.

\vspace{.12in} \noindent \textbf{2.4. Systems with configuration
space; localizability}

\vspace{.12in} We shall now consider the class of systems each of
which has a configuration space (say, M) associated with it and it
is meaningful to ask questions about the localization of the system
in subsets of M. To start with, we shall take M to be a topological
space and take the permitted domains of localization to belong to
B(M), the family of Borel subsets of M.

Some good references containing detailed treatment of localization
in conventional approaches are (Newton and Wigner [37], Wightman
[46], Varadarajan [44], Bacry [4]). We shall follow a relatively
more economical path exploiting some of the constructions described
above and in I.

We shall say that a system S [with associated symplectic
superalgebra $(\sca,\omega)$] is \emph{localizable} in M if we have
a positive observable-valued measure (as defined in section 2.1
above) on the measurable space (M,B(M)), which means that,
corresponding to every subset $ D \in B(M)$, there is a positive
observable P(D) in \sca \ satisfying  the three conditions

\noindent (i) $ P(\emptyset) = 0;$ \hspace{.5in} (ii) P(M) = I;

\noindent (iii) for any countable family of mutually disjoint sets
$D_i \in B(M)$,
\begin{eqnarray}
P (\cup_iD_i) = \sum_i P(D_i).
\end{eqnarray}
For such a system, we can associate, with any state $\phi$, a
probability measure $\mu_{\phi}$ on the measurable space $(M,B(M))$
defined by [see Eq.(1)]
\begin{eqnarray}
\mu_{\phi}(D) = \phi (P(D)),
\end{eqnarray}
making the triple $(M, B(M), \mu_{\phi})$ a probability space. The
quantity $\mu_{\phi}(D)$ is to be interpreted as the probability of
the system, given  in the state $\phi$, being found (on
observation/measurement) in the domain D.

Generally it is of  interest to consider localizations having
suitable invariance properties under a transformation group G.
Typically G is a topological group  with continuous action on M
assigning, to each $g \in G$, a bijection $T_g : M \rightarrow M$
such that, in obvious notation, $T_g T_{g^{\prime}} =
T_{gg^{\prime}}$ and $T_e = id_M$; it also has a symplectic action
on \sca \ and $\scs(\sca)$ given by  the mappings $\Phi_1(g)$ and
$\Phi_2(g)$ introduced in section 3.5 of I [$\Phi_1(g)$, for every
$g \in G$,  is a canonical transformation of \sca \ and $\Phi_2(g)$
= $([\Phi_1(g)]^{-1})^T$ acts on states].

. The localization in M described above will be called
\emph{G-covariant} (or, loosely, G-invariant) if \begin{eqnarray}
\Phi_1(g)(P(D)) = P(T_g(D)) \ \ \forall g \in G \ \textnormal{and} \
D \in B(M).
\end{eqnarray}

\vspace{.1in} \noindent \textbf{Proposition 2.2} \emph{In a
G-covariant localization as described above, the localization
probabilities (7) satisfy the covariance condition}
\begin{eqnarray} \mu_{\Phi_2(g)(\phi)}(D) =
\mu_{\phi}(T_{g^{-1}}(D)) \ \textnormal{for all} \ \phi \in
\scs(\sca) \ \textnormal{and} \ D \in B(M).
\end{eqnarray}
\emph{Proof.} We have \begin{eqnarray*} \mu_{\Phi_2(g)(\phi)}(D) & =
& <\Phi_2(g)(\phi), P(D)> \ = \ <\phi, \Phi_1(g^{-1})(P(D))> \\ & =
& <\phi, P(T_{g^{-1}}(D))> \ = \ \mu_{\phi}(T_{g^{-1}}(D)). \ \ \Box
\end{eqnarray*}

In most practical applications, M is a manifold and G  a Lie group
with smooth action on M and a Poisson action on the symplectic
superalgebra $(\sca, \omega)$. In this case, the `hamiltonian'
$h_{\xi}$ corresponding to an element $\xi$ of the Lie algebra \scg
\ of G is an  observables which serves, through Poisson brackets, as
the infinitesimal generator of the one-parameter group of  canonical
transformations induced by the action of the one-parameter group
generated by $\xi$ on the system algebra \sca \ (I, section 3.5).
The Poisson brackets between these hamiltonins correspond to the
commutation relations in \scg \ [se Eq.(59) in I and Eq.(13) below].

In Hilbert space QM, the problem of G-covariant localization is
traditionally formulated in terms of the so-called `systems of
imprimitivity' (Mackey [34], Varadarajan [44], Wightman [46]). We
are operating in the more general algebraic setting trying to
exploit the machinery of noncommutative symplectic geometry
developed in I.
Clearly, there is considerable scope for mathematical developments
in this context parallel to those relating to systems of
imprimitivity. We shall, however, restrict ourselves to some
essential developments relevant to the treatment of localizable
elementary systems (massive particles) later.

We shall be mostly concerned with $ M= \mathbb{R}^n$ (equipped with
the Euclidean metric). In this case, one can consider averages of
the form (denoting the natural coordinates on $\mathbb{R}^n$ by
$x_j$)
\begin{eqnarray}
\int_{\mathbb{R}^n} x_j d \mu_{\phi}(x), \ \ j= 1,...,n.
\end{eqnarray}
It is natural to introduce \emph{position/configuration observables}
$X_j$ such that the quantity (10) is $\phi(X_j)$. Let $E_n$ denote
the (identity component of) Euclidean group in n dimensions and let
$p_j, m_{jk} (= - m_{kj})$ be its generators satisfying the
commutation relations \begin{eqnarray}  [p_j,p_k] = 0, \ \ [m_{jk},
p_l] = \delta_{jl}p_k - \delta_{kl}p_j  \nonumber \\  {[}m_{jk},
m_{pq}] = \delta_{jp}m_{kq} - \delta_{kp}m_{jq} - \delta_{jq}m_{kp}
+ \delta_{kq}m_{jp}. \end{eqnarray} We shall say that a system S
with configuration space $\mathbb{R}^n$ has \emph{concrete
Euclidean-covariant localization} if it is localizable as above in $
\mathbb{R}^n$ and

\noindent (i) it has position observables $X_j \in \sca$ \ such
that, in any state $\phi$,
\begin{eqnarray}
\phi(X_j) & = & \int_{R^n} x_j d \mu_{\phi}(x);
\end{eqnarray}
(The term `concrete' is understood to imply this condition.)

\noindent (ii) the group $E_n$ has a Poisson action on \sca \ so
that we have the hamiltonians $P_j, \ M_{jk}$ associated with the
generators $p_j, m_{jk}$ such that
\begin{eqnarray}
\{ P_j, P_k \} = 0, \hspace{.12in}  \{ M_{jk}, P_l \} =
\delta_{jl}P_k -
\delta_{kl} P_j \nonumber \\
\{ M_{jk}, M_{pq} \} = \delta_{jp} M_{kq} - \delta_{kp}M_{jq} -
\delta_{jq} M_{kp} + \delta_{kq} M_{jp};
\end{eqnarray}

\noindent (iii) the covariance condition (9) holds with the
Euclidean group action on $\mathbb{R}^n$ given by
\begin{eqnarray} T_{(R,a)} x = Rx + a, \ \ R \in SO(n), \ a \in \mathbb{R}^n.
\end{eqnarray}

\vspace{.1in} \noindent \textbf{Proposition 2.3} \emph{For supmech
systems with concrete Euclidean - covariant localization in
$\mathbb{R}^n$, the infinitesimal Euclidean transformations of the
localization observables $X_j$ are given by the PB relations}
\begin{eqnarray}
\{ P_j, X_k \} = \delta_{jk} I , \hspace{.12in} \{M_{jk}, X_l \} =
\delta_{jl} X_k - \delta_{kl} X_j.
\end{eqnarray}
\emph{Proof}. Using Eq.(12) with $\phi$ replaced by $\phi^{\prime} =
\Phi_2(g)(\phi)$, we have
\begin{eqnarray*}
\phi^{\prime}(X_j) = \int x_j d\mu_{\phi^{\prime}}(x) = \int x_j d
\mu_{\phi}(x^{\prime}) = \int (x_j^{\prime} - \delta x_j) d
\mu_{\phi}(x^{\prime}) \end{eqnarray*} where $x^{\prime} \equiv
T_{g^{-1}}(x) \equiv x + \delta x$ and we have used Eq.(9) to write
$d \mu_{\phi^{\prime}}(x) = d \mu_{\phi}(x^{\prime})$. [Application
of the transformation rule for integration over a measure
(DeWitt-Morette and Elworthy [19]; p.130) gives the same result.]
Writing $\phi^{\prime} = \phi + \delta \phi$ and taking $T_g$ to be
a general infinitesimal transformation generated by $ \epsilon \xi =
\epsilon^a \xi_a$, we have [recalling Eq.(53) of I]
\begin{eqnarray} - (\delta \phi)(X_j) =
\epsilon \phi ( \{ h_{\xi}, X_j \}) = \int_{R^n} \delta x_j d
\mu_{\phi}(x). \end{eqnarray} For translations, with $ \xi = p_k, \
h_{p_k}= P_k, \ \delta x_j = \epsilon \delta_{jk}$, Eq.(16) gives
\begin{eqnarray*} \phi(\{ P_k, X_j \}) = \delta_{jk} = \delta_{jk}
\phi(I). \end{eqnarray*} Since this holds for all $\phi \in \scs
(\sca)$, we have the first of the equations (15). The second
equation is similarly obtained by taking, in obvious notation, $
\epsilon \xi = \frac{1}{2}\epsilon_{jk} m_{jk}$ and
\begin{eqnarray*} \delta x_l = \epsilon _{lk}x_k = \epsilon_{jk}
\delta_{ jl} x_k = \frac{1}{2} \epsilon_{jk} (\delta_{jl}x_k -
\delta_{kl} x_j). \  \Box \end{eqnarray*}

 The hamiltonians $ P_{j} $ and $ M_{jk} $ will be referred to as the
\emph{momentum} and \emph{angular momentum} observables of the
system S. It should be noted that the PBs obtained above do not
include the expected relations $ \{ X_j, X_k \} =0$; these
relations, as we shall see in the following subsection, come from
the relativity group. [Recall that, in the treatments of
localalization based on systems of imprimitivity, the commutators
$[X_j,X_k] = 0$ appear because there the analogues of the objects
P(D) are assumed to be projection operators satisfying the relation
$P(D)P(D^{\prime}) = P(D \cap D^{\prime})  (= P(D^{\prime})P(D))$.
In our more general approach, we do not have such a relation.]

\vspace{.15in} \noindent \textbf{2.5. Elementary systems; Particles}

\vspace{.12in} We shall now obtain, in the framework of supmech, the
fundamental observables relating to the characterization/labelling
and kinematics of a particle. Relativity group will be seen to play
an important role in this context.

 Particles are irreducible entities localized in `space' and their
dynamics involves `time'. Their description, therefore, belongs to
the subdomain of supmech admitting  space-time descriptions of
systems. The space-time M will be assumed here to be a (3+1)-
dimensional differentiable manifold equipped with a suitable metric
to define spatial distances and time-intervals. A \emph{reference
frame} is an atlas on M providing a coordinatization of its points.
\emph{Observers} are supposedly intelligent beings employing
reference frames for doing concrete physics; they will be understood
to be in one-to-one correspondence with reference frames.

To take into consideration observer-dependence of observables, we
adopt the \emph{principle of relativity} formalized as follows :

\vspace{.1in} \noindent (i) There is a preferred class of reference
frames whose space-time coordinatisations are related through the
action of a connected Lie group $G_0$ (the \emph{relativity group}).

\noindent (ii) The relativity group $G_0$ has a hamiltonian action
on the symplectic superalgebra $(\sca, \omega)$ [or the generalized
symplectic superalgebra $(\sca, \scx, \omega)$ (see section 3.7 of I)
in appropriate situations] associated with a system.

\noindent (iii) All reference frames in the chosen class are
physically equivalent in the sense that the fundamental equations of
the theory are covariant with respect to the $G_0$-transformations
of the relevant variables.

\vspace{.1in} We shall call such a scheme \emph{$G_0$-relativity}
and systems covered by it \emph{$G_0$-relativistic}. In the present
work, $G_0$ will be assumed to have the one-parameter group
$\mathcal{T}$ of time translations as a subgroup. This allows us to
relate the Heisenberg and Schr$\ddot{o}$dinger pictures of dynamics
corresponding to two observers O and O$^{\prime}$ through the
symplectic action of $G_0$ by following the strategy adopted in
(Sudarshan and Mukunda  [43]; referred to as SM below). Showing the
observer dependence of the algebra elements explicitly, the two
Heisenberg picture descriptions A(O,t) and
A(O$^{\prime}$,t$^{\prime}$) of an element A of \sca \ can be
related through the sequence (assuming a common zero of time for the
two observers)
\[ A(O,t) \longrightarrow A(O,0) \longrightarrow A(O^{\prime},0)
\longrightarrow A(O^{\prime},t^{\prime}) \] where the first and the
last steps involve the operations of time translations in the two
frames. We shall be concerned only with the symplectic action of
$G_0$ on \sca \ involved in the middle step.

To formalize the notion of a (relativistic, quantum) particle as an
irreducible entity, Wigner [48] introduced the concept of an
`elementary system' as a quantum system whose Hilbert space carries
a projective unitary irreducible representation of the
Poincar$\acute{e}$ group. The basic idea is that the state space of
an elementary system should not admit a decomposition into more than
one invariant (under the action of the relevant relativity group)
subspaces. Following this idea, elementary systems in classical
mechanics (SM; Alonso [2]) have been defined in terms of a
transitive action of the relativity group on the phase space of the
system. Our treatment of elementary systems in supmech will cover
classical and quantum elementary systems  as special cases.

A system S having associated with it the symplectic triple $(\sca,
\sone, \omega)$ will be called an \emph{elementary system} in
$G_0$-relativity  if it is a $G_0$-relativistic system such that the
action of $G_0$ on the space \sone \ of its pure states is
transitive. Formally, an elementary system may be represented as a
collection $ \mathcal{E} = ({G}_0, \sca, \sone,\omega, \Phi)$ where
$ \Phi = (\Phi_1, \Phi_2)$ are mappings as in section 3.5 of I
implementing the $G_0$-actions --- $\Phi_1$ describing a hamiltonian
action on $(\sca, \omega)$ and $\Phi_2 [= (\tilde{\Phi}^{-1})]$ a
transitive action on \sone.

\vspace{.1in} \noindent \textbf{Proposition 2.4} \emph{In the
$G_0$-relativity scheme, a  $G_0$-invariant observable must be a
multiple of the unit element.}

\vspace{.1in} \noindent \emph{Proof.} Let Q be such an observable
and $\phi_1, \phi_2$ two pure states. The transitive action of $G_0$
on \sone \ implies that $\phi_2 = \Phi_2(g) (\phi_1) $ for some $ g
\in G_0$. We have
\[ <\phi_2, Q> \ = \ <\Phi_2(g)(\phi_1), Q>\  = \ <\phi_1, \Phi_1(g^{-1})(Q)>
= <\phi_1, Q> \] showing that the expectation value of Q is the same
in every pure state. Denoting this common expectation value of Q by
q (we shall call it the \emph{value} of Q for the system), we have,
by the CC condition, Q = qI. $ \Box$

This has the important implication that, for an elementary system, a
Poisson action [of $G_0$ or of its projective group $\hat{G}_0$ (see
section 3.5 of I)] is always available; this is because, if $G_0$
does not admit Poisson action, the values $\alpha(\xi, \eta)$ of the
cocycle $\alpha$ of section 3.5 of I (where $\xi, \eta$ are elements
of the Lie algebra $\scg_0$ of $G_0$), since they have vanishing
Poisson brackets (PBs) with all elements of \sca \ (hence with the
hamiltonians corresponding to $G_0$), are multiples of the unit
element and the hamiltonian action of $G_0$ can be extended to a
Poisson action of $\hat{G}_0$. [See the discussion following Eq.(62)
of I.] In the remainder of this subsection, $\hat{G}_0$ will stand
for the effective relativity group which will be $G_0$ or its
projective group depending on whether or not $G_0$ admits Poisson
action on \sca.

Let $\xi_a$ (a = 1,..,r) be a basis in the Lie algebra
$\hat{\scg}_0$ of $\hat{G}_0$ satisfying the commutation relations $
[ \xi_a, \xi_b] = C_{ab}^c \xi_c.$  Corresponding to the generators
$\xi_a$, we have the hamiltonians $h_a \equiv h_{\xi_a}$ in \sca \
satisfying the  PB relations
\begin{eqnarray}
\{ h_a, h_b \} = C_{ab}^c \ h_c.
\end{eqnarray}
These relations are the same for all elementary systems in
$G_0$-relativity.

In classical mechanics, one has an isomorphism between the
symplectic structure on the symplectic manifold of an elementary
system and that on a coadjoint orbit in $\hat{\scg}_0^*$ (the
conjugate space of the Lie algebra $\hat{\scg}_0$). In our case, the
state spaces of elementary systems and coadjoint orbits of
relativity groups are generally spaces of different types and the
question of an isomorphism does not arise. The appropriate relation
in supmech corresponding to the above mentioned relation in
classical Hamiltonian mechanics is given by proposition 2.5 below.
Adopting/(adapting from) the notation of section 3.6 of I, we have
the mapping $ h: \hat{\scg}_0 \rightarrow \sca $ given by $h(\xi) =
h_{\xi}$; the noncommutative momentum map is the restriction to
\sone \ of the transposed map $\tilde{h} : \sca^* \rightarrow
\hat{\scg}_0^*$ :
\begin{eqnarray}
<\tilde{h}(\phi), \xi> \ = \ < \phi, h(\xi) > \ = \ < \phi, h_{\xi}>
\ \ \textnormal{for all}\  \phi \in \sone.
\end{eqnarray}
The equivariance condition for the noncommutative momentum map
$\tilde{h}$ [(68) of I] is
\begin{eqnarray} \tilde{h}(\Phi_2(g) \phi)
= Cad_g(\tilde{h}(\phi))
\end{eqnarray}
where Cad stands for the co-adjoint action of $\hat{G}_0$ on
$\hat{\scg}_0^*$.

\vspace{.1in} \noindent \textbf{Proposition 2.5} \emph{Adopting the
notations introduced above in the context of elementary systems in
$G_0$-relativity, we have \\
(a) the $\tilde{h}$-images of pure states of an elementary system in
supmech are co-adjoint orbits; \\
(b) the coordinates $u_a(g)$ of a general point of the co-adjoint
orbit corresponding to the pure state $\phi$ [defined by
$Cad_g[\tilde{h}(\phi)] = u_a(g) \lambda^a $ where $\{ \lambda^a \}$
is the dual basis in $\hat{\scg}_0^*$ corresponding to the basis $\{
\xi_a \}$ in $\hat{\scg}_0$] are given by}
\begin{eqnarray}
u_a(g) = <\phi, \Phi_1(g^{-1})h_a>.
\end{eqnarray}
\emph{Proof.} Part (a) follows immediately from Eq.(19) and the
transitivity of the $\hat{G}_0$-action on the pure states.

\noindent Part(b). We have
\begin{eqnarray*}
u_a(g) = \ < Cad_g [\tilde{h}(\phi)], \xi_a> \ = \
<\tilde{h}[\Phi_2(g)(\phi)], \xi_a> \  = \ <\phi,
\Phi_1(g^{-1})h_a>. \ \Box
\end{eqnarray*}
Eq.(20) shows that the transformation properties of the hamiltonians
$h_a$ are directly related to those of the corresponding coordinates
(with respect to the dual basis) of points on the relevant
co-adjoint orbit. This is adequate to enable us to to use the
descriptions of the relevant co-adjoint actions in (Alonso [2]) and
draw parallel conclusions.

For the treatment of elementary systems in a given relativity
scheme, we shall adopt the following strategy :

\vspace{.1in} \noindent (i) Obtain the PBs (17).

\vspace{.1in} \noindent (ii) Use these PBs to identify
\emph{fundamental observables} [i.e. those which cannot be obtained
from other observables (through algebraic relations or PBs)]. These
include observables (like mass) that Poisson-commute with all $h_a$s
and the momentum observables (if the group of space translations is
a subgroup of the relativity group  considered).

\vspace{.1in} \noindent (iii) Determine the transformation laws of
$h_a$s under finite transformations of $G_0$ following the relevant
developments in (SM; Alonso [1]). Use these transformation laws to
identify the $G_0$-invariants and some other fundamental observables
(the latter are configuration and spin observables in the schemes of
Galilean and special relativity). The values of the invariant
observables serve to characterize/label an elementary system.

\vspace{.1in} \noindent (iv) The system algebra \sca \ for an
elementary system is to be taken as the one generated by the
fundamental observables and the identity element.

\vspace{.1in} \noindent (v) Obtain (to the extent possible) the
general form of the Hamiltonian as a function of the fundamental
observables as dictated by the PB relations (17).

For illustration, we consider the scheme of Galilean relativity.

\vspace{.12in} \noindent \textbf{Nonrelativistic elementary systems}

\vspace{.1in} In the nonrelativistic domain, the relativity group
$G_0$ is the Galilean group of transformations of the Newtonian
space-time $\mathbb{R}^3 \times \mathbb{R}$ given by
\begin{eqnarray}
g = (b,a,v,R) : (x,t) \mapsto ( Rx + tv + a, t + b)
\end{eqnarray}
where $R \in SO(3),\  v \in \mathbb{R}^3, \ a \in \mathbb{R}^3$ and
$b \in \mathbb{R}$. Choosing a basis of the 10-dimensional Lie
algebra $\scg_0$ of $G_0$ in accordance with the representation
\[ g = exp(b\sch) \ exp (a.\mathcal{P})\  exp (v.\mathcal{K})\  exp
(w.\mathcal{J}) \] most of the the commutators among the
generators $ \mathcal{J}_j, \mathcal{K}_j, \mathcal{P}_j, \mathcal{H}$
are standard or obvious; the nontrivial commutators are
\begin{eqnarray} [\mathcal{K}_j, \sch] = \mathcal{P}_j, \ \
[\mathcal{K}_j, \mathcal{P}_k] = 0. \end{eqnarray} [In fact, the last one
should also be obvious from Eq.(21); it has been recorded here for its
special role below.]

Recalling the discussion relating to Poisson action of Lie groups on
symplectic superalgebras in section 3.5 of I, the cohomology group $
H_0^2(\scg_0, \mathbb{R})$ does not vanish (implying
non-implementability of a Poisson action of $G_0$) and has dimension
one (Cari$\tilde{n}$ena,  Santander [13]; Alonso [2]; Guillemin,
Sternberg [25]; SM). Choosing the representative cocycle in
$Z_0^2(\scg_0, \mathbb{R})$ as $ \eta(\mathcal{K}_j, \mathcal{P}_k)
= - \delta_{jk} \mathcal{M}$ (where $\mathcal{M}$ is the additional
generator), Eq. (63) of I implies the replacement
of the second equation in (22) by
\begin{eqnarray} [\mathcal{K}_j, \mathcal{P}_k] = -\delta_{jk}
\mathcal{M}. \end{eqnarray} Supplementing the so modified
commutation relations of $\scg_0$ with the vanishing commutators of
$\mathcal{M}$ with the ten generators of $\scg_0$, we obtain the
commutation relations of the 11-dimensional Lie algebra
$\hat{\scg}_0$ of the projective group $\hat{G}_0$ of the Galilean
group $G_0$.

 The hamiltonians $ J_i, K_i, P_i, H, M $  corresponding to the
generators $\mathcal{J}_i, \mathcal{K}_i, \linebreak \mathcal{P}_i
(i=1,2,3), \sch, \mathcal{M}$ of $\hat{G}_0$  [so that
$h_{\mathcal{P}_i} = P_i$ etc] satisfy the Poisson bracket relations
(SM)
\begin{eqnarray}
\{J_i, J_j \} = - \epsilon_{ijk}J_k, \ \ \{J_i, K_j \} = -
\epsilon_{ijk} K_k,
\ \  \{ J_i, P_j \} = - \epsilon_{ijk} P_k  \nonumber \\
 \{ K_i, H \} = - P_i, \ \ \{ K_i, P_j \} = - \delta_{ij} M;
\end{eqnarray}
all other PBs vanish. By the argument presented above, we must have
M= mI, $m \in \mathbb{R}$. We shall identify m as the mass of the
elementary system. The condition $m \geq 0$ will follow later from
an appropriate physical requirement. The objects $P_i$ and $J_i$,
being generators of the Euclidean subgroup $ E_3 $ of $G_0$, are the
momentum and angular momentum observables of subsection 2.4 above.

The transformation laws of the hamiltonians of $\hat{G}_0$ under its
adjoint action (SM;  Alonso [2]) yield  the following three
independent invariants
\begin{eqnarray}
M, \hspace{.12in} C_1 \equiv 2MH - \mathbf{P}^2, \hspace{.12in} C_2
\equiv (M \mathbf{J} - \mathbf{K} \times \mathbf{P})^2.
\end{eqnarray}
Of these, the first one is obvious; the vanishing of PBs of $C_1$
with all the hamiltonians is also easily checked. Writing $C_2 =
B_jB_j$ where
\begin{eqnarray*} B_j = MJ_j - \epsilon_{jkl}K_kP_l, \end{eqnarray*}
it is easily verified that
\begin{eqnarray*} \{J_j, B_k \} = - \epsilon_{jkl} B_l, \ \
\{K_j, B_k \} = \{ P_j, B_k \} = \{H, B_k \} = 0 \end{eqnarray*}
which finally leads to the vanishing of PBs of $C_2$ with all the
hamiltonians. The values of these three invariants characterize a
Galilean elementary system in supmech.

We henceforth restrict ourselves to elementary systems with $ m \neq
0.$ Defining $X_i = m^{-1} K_i$, we have
\begin{eqnarray}
\{ X_j, X_k \} = 0, \hspace{.12in} \{ P_j, X_k \} = \delta_{jk} I, \
\ \hspace{.12in} \{ J_j, X_k \} = - \epsilon_{jkl} X_l.
\end{eqnarray}
 Comparing the last two equations above with the equations (15)(for n=3),
we identify $X_j$  with the position observables of section 2.4.
Note that, as mentioned earlier, the fact that the $X_j$s mutually
Poisson-commute comes from the relativity group.

Writing $ \mathbf{S} = \mathbf{J} - \mathbf{X} \times \mathbf{P}, $
we have  $C_2$ = $m^2 \mathbf{S}^2$. We have the PB relations
\begin{eqnarray}
\{ S_i, S_j \} = - \epsilon_{ijk} S_k, \ \  \{ S_i, X_j \} = 0 = \{
S_i, P_j \}.
\end{eqnarray}
We identify $\mathbf{S}$ with the internal angular momentum or spin
of the elementary system.

The invariant quantity
\begin{eqnarray}
U \equiv \frac{C_1}{2m} = H - \frac{\mathbf{P}^2}{2m}
\end{eqnarray}
is interpreted as the \emph{internal energy} of the elementary
system; its appearance as one of the invariant observables  of a
Galilean elementary system reflects the possibility that such an
elementary system may have an internal dynamics involving dynamical
variables which are invariant under the action of the Galilean
group. It is the appearance of this quantity (which plays no role in
Newtonian mechanics) which is responsible for energy being defined
in Newtonian mechanics only up to an additive constant.

Writing $\mathbf{S}^2 = \sigma I$ and U = u I, we see that Galilean
elementary systems with $m \neq 0$ can be taken to be
characterized/labelled  by the parameters m, $\sigma$ and u. The
fundamental kinematical observables are $X_j, P_j$ and $S_j$
(j=1,2,3). The system algebra \sca \ of a nonrelativistic elementary
system is assumed to be the one generated by the fundamental
observables and the identity element.

Particles are defined as the elementary systems with  u = 0. Eq.(28)
now gives
\begin{eqnarray}
H = \frac{\mathbf{P}^2}{2m}
\end{eqnarray}
which is the Hamiltonian for a free Galilean particle in supmech.

\vspace{.1in} \noindent Note. (i) Full Galilean invariance (more
generally, full invariance under a relativity group) applies only to
an isolated system. Interactions/(external influences) are usually
described with (explicit or implicit) reference to a fixed reference
frame or a restricted class of frames. For example, the interaction
described by a central potential implicitly assumes that the center
of force is at the origin of axes of the chosen reference frame.

\noindent (ii) In the presence of external influences, invariance
under space translations is lost and the PB $ \{H, P_i \} = 0 $ must
be dropped. For a spinless particle, the Hamiltonian, being an
element of the system algebra generated by the fundamental
observables \textbf{X} and \textbf{P} and the unit element I,
has the general form
\begin{eqnarray}
H = \frac{\mathbf{P}^2}{2m} + V(\mathbf{X}, \mathbf{P}).
\end{eqnarray}
In most practical situations, V is a function of $ \mathbf{X}$ only.

\vspace{.1in} The Hamiltonian was assumed in section 3.4 of I to be
bounded below (in the sense that its expectation values in all
states are bounded below); this rules out the case $m < 0$ because,
by Eq.(29), this will allow arbitrarily large negative expectation
values for energy. (Expectation values of the observable $
\mathbf{P}^2$ are expected to have no upper bound.)

Recalling the demonstration of the classical Hamiltonian mechanics
as a subdiscipline of NHM in section 3.4 of I, the classical
Hamiltonian system for a  massive spinless Galilean particle is
easily seen to be the special case of the corresponding supmech
Hamiltonian system with $\sca = C^{\infty}(\mathbb{R}^6)$. The
corresponding quantum system is also (recalling the example in
section 3.3 of I) a special case of a supmech Hamiltonian system
with the system algebra generated by the position and momentum
observables in Schr$\ddot{o}$dinger theory. More detailed treatment
(with justification of the Schr$\ddot{o}$dinger theory) will appear
in section 3.4.

\vspace{.12in} \noindent \textbf{2.6. Noncommutative  Noether
invariants of the projective Galilean group for a free massive spinless
particle}

\vspace{.1in} In section 3.9 of I, the noncommutative analogue of
the symplectic version of Noether's theorem was proved. Given a
symplectic superalgebra $(\sca, \omega)$ and a Hamiltonian H as an
element of  the extended system algebra $ \sca^e =
C^{\infty}(\mathbb{R}) \otimes \sca $, one constructs a
presymplectic algebra $(\sca^e, \Omega)$ where
\[ \Omega = \tilde{\omega} - dH \wedge dt. \] Here the real
line $\mathbb{R}$ is the carrier space of the evolution parameter
(`time') t, $\tilde{\omega} = 1 \otimes \omega$ is the isomorphic
copy of the symplectic form $\omega$ in the subalgebra $\tilde{\sca}
= 1 \otimes \sca$ of $\sca^e$ and d is the exterior derivative in
the differential calculus based on $\sca^e$ induced by the exterior
derivatives $d_1$ and $d_2$  in the differential calculi based on
the algebras $C^{\infty}(\mathbb{R})$ and \sca \ respectively,
according to Eq.(28) of I [giving, on identifying t with $t \otimes
I$, where I is the unit element of \sca, $dt = d_1 t \otimes I$].

A canonical transformation on the presymplectic superalgebra
$(\sca^e, \Omega)$, was defined in I as a superalgebra isomorphism
$ \Phi : \sca^e \rightarrow \sca^e$ such
that (i) $ \Phi^* \Omega = \Omega$, (ii) $\Phi (\tilde{\sca}_0 )
\subset \tilde{\sca}_0$ where $\sca_0 \equiv C^{\infty}(\mathbb{R})$
and $\tilde{\sca}_0 = \sca_0 \otimes I \subset \sca^e$. This is in
keeping with the tradition that the Noetherian symmetries map, in
particle mechanics, the `time' space into itself and, in field
theory, space-time into itself.

When a Lie group G with Lie algebra \scg \ has a symplectic action
on this presymplectic algebra,  the induced infinitesimal generator
$\hat{Z}_{\xi}$ corresponding to $\xi \in \scg$ satisfies the condition
$L_{\hat{Z}_{\xi}} \Omega = 0$ which, with $d \Omega = 0$, implies
\[ d ( i_{\hat{Z}_{\xi}} \Omega ) = 0. \]
When the G-action is hamiltonian, we have [see Eq.(76) of I]
\begin{eqnarray}
i_{\hat{Z}_{\xi}} \Omega = -d \hat{h}_{\xi};
\end{eqnarray}
in this case, the noncommutative symplectic Noether's theorem
[theorem (1) in I] states that the `hamiltonians' $ \hat{h}_{\xi}$
are constants of motion.

\vspace{.1in} \noindent \emph{Note.} The traditional Noether's
theorem has its development in the classical Lagrangian formalism.
It has an equivalent in the `time dependent' Hamiltonian formalism
[1] based on the presymplectic manifold $(\mathbb{R} \times T^*M,
\tilde{\omega}_0)$ where M is the configuration manifold and
$\tilde{\omega}_0$ is the pull-back on $\mathbb{R} \times T^*M$ of
the canonical symplectic form $\omega_0$ on the cotangent
bundle $T^*M$. This
symplectic version admits a generalization ([42], I) to more general
presymplectic manifolds. The theorem proved in I is the
noncommutative analogue of this more general symplectic version of
Noether's theorem (restricted to the class of presymplectic
manifolds obtained by replacing $T^*M$ above by a general symplectic
manifold P). In particular, Eq.(31) is the noncommutative analogue
of the equation in Def.(11.7b) on p. 101 of [42].

\vspace{.1in} Here we are interested in the explicit construction of
the Noether invariants $\hat{h}_{\xi}$ when G is the projective
group $\hat{G}_0$ of the Galilean group $G_0$ and \sca \ the algebra
generated by the fundamental observables $X_j$ and $P_j$ (j=1,2,3)
of a free nonrelativistic spinless particle and the identity element
I and H is given by Eq.(29). Construction of these objects involves
consideration of the transformation of the time variable which was
bypassed in the previous subsection. Some caution is needed in the
treatment of time translations. In this case, we have $\hat{Z}_{\xi} =
\frac{\partial}{\partial t}$ [ and not $ \hat{Z}_{\xi} =
\frac{\partial}{\partial t} + \{ H,. \}$]. The point is that, stated
in very general terms, Noether's theorem says that, given an
invariance property of a certain object, we have a conserved
quantity in dynamics. For the consideration of invariance of
the relevant object (`action' in the classical Lagrangian
formalism and $\Omega$ in the present context), one employs the
kinematical transformations (corresponding to the group action)
on the relevant variables; for this the induced derivation for
time translations is $ \frac{\partial}{\partial t}$ [see, for
example, Eq.(55) in Ch.(6) of Dass [14] which has, for the
action of an infinitesimal time translation in the context of a
system of interacting particles, $ \delta t = \epsilon, \ 
\delta \mathbf{r}_A = 0$ in obvious notation]. On the other hand,
while checking for conservation of a quantity, one considers
change in the  quantity when the system point moves on a
dynamical trajectory; for this, the appropriate derivation is, of
course, $\frac{\partial}{\partial t} + \{ H,. \}.$  [Recall the statement in
mechanics : `If the Lagrangian has no \emph{explicit} dependence on
time, then the Hamiltonian/energy is a constant of motion.']

For $\xi \in \hat{\scg}_0$ corresponding to other transformations
( $\xi = \mathcal{J}_i, \mathcal{P}_i, \mathcal{K}_i, \mathcal{M}$ ), we have
the usual Poisson action of $\hat{G}_0$ on \sca \ (identified with
$\tilde{\sca} = 1 \otimes \sca$) for which $\hat{Z}_{\xi} = Y_{h_{\xi}}$
where $h_{\xi}$ is the hamiltonian corresponding to $\xi$.

Recalling the notation introduced in section 3.8 of I, we have, for
any element $ F = \sum_i f_i \otimes F_i$ of $\sca^e$,
\begin{eqnarray} dF = \sum_i (d_1 f_i) \otimes F_i + \sum_i f_i \otimes (d_2 F_i)
\equiv \tilde{d}_1 F + \tilde{d}_2 F. \end{eqnarray}
 Note that $\tilde{d}_1 F =
\frac{\partial F}{\partial t} dt;$ hence $dt \wedge d H = dt \wedge
\tilde{d}_2 H.$

Equation (32) and the equation defining $\Omega$ above now give
\begin{eqnarray}
i_{Y_{h_{\xi}}} \Omega & = & i_{Y_{h_{\xi}}} \tilde{\omega} -
i_{Y_{h_{\xi}}}(\tilde{d}_2 H) dt  \nonumber \\
              & = & -\tilde{d}_2 h_{\xi} - \{h_{\xi},H \} dt.
\end{eqnarray}
In the calculations presented below, the various hamiltonians
$h_{\xi}$  have no explicit time dependence; hence, in the last
line in Eq.(33), we have $\tilde{d}_2 h_{\xi} = d h_{\xi}$. 

Coming back to Eq.(31), we now have

\vspace{.1in} \noindent (i) for rotations ($ \xi = \mathcal{J}_i,
h_{\xi} = J_i $),  $\{h_{\xi},H \} = 0$,  giving $
\hat{h}_{\xi} = h_{\xi} = J_i; $

\vspace{.1in} \noindent (ii) for space translations ( $ \xi =
\mathcal{P}_i, h_{\xi} = P_i $),  $\{h_{\xi},H \} = 0$,
giving $\hat{h}_{\xi} = h_{\xi} = P_i;$

\vspace{.1in} \noindent (iii) for Galilean boosts ($ \xi =
\mathcal{K}_i, h_{\xi} = K_i = m X_i $),  
 $ \{K_i, H \} = -  P_i$ giving $\hat{h}_{\xi} = m X_i
- P_i t$;

\vspace{.1in} \noindent (iv) for time translations,
$\hat{Z}_{\xi} = \frac{\partial}{\partial t}, i_{\hat{Z}_{\xi}}
\Omega = dH$, giving $\hat{h}_{\xi}= - H$;

\vspace{.1in} \noindent (v) for the one-parameter group generated by
$\mathcal{M}$ ($\xi = \mathcal{M}, \ h_{\xi} = M = mI$),
$\{h_{\xi},H \} = 0$, giving $\hat{h}_{\xi}= M = m I$.

\vspace{.1in} \noindent Finally, we have

\vspace{.1in} \noindent \textbf{Proposition 2.6} \emph{ The
noncommutative Noether invariants of the projective group
$\hat{G}_0$ of the Galilean group $G_0$ for a free nonrelativistic
spinless particle of mass m are}
\begin{eqnarray}
\mathbf{J}, \ \mathbf{P},\ m \mathbf{X} - \mathbf{P}t , \ \  - H, \
\ M = mI.
\end{eqnarray}
Note that the first four of these  are (up-to signs) the supmech
avatars of those in (Souriau [42]; p.162).

\vspace{.1in} \noindent \emph{Note.} If, instead of taking $X_j =
m^{-1} K_j$ in a treatment bypassing the involvement of time in the
symplectic transformations  as above, we had proceeded to identify
observables through Noether invariants, we would  have got the
position observable as $m^{-1}$ times the time-independent term in
the third entry in the list (34).

\vspace{.15in} \noindent \textbf{3. Quantum Systems}

\vspace{.1in}  We now take up a systematic study of the  `quantum
systems' defined as supmech Hamiltonian systems with
non-supercommutative system algebras. Theorem (2) of I dictates
these systems to have a standard symplectic structure characterized
by a universal real parameter of the dimension of action; we shall
identify it with the Planck constant $\hbar$. We first treat quantum
systems in the general algebraic setting. We then employ the CC
condition to show that they inevitably have Hilbert space based
realizations, generally admitting commutative superselection rules.
The autonomous development of the Hilbert space QM of `standard
quantum systems' (those with finitely generated system algebras) is
then presented. This is followed by a straightforward treatment of
the Hilbert space quantum mechanics of material particles.

\vspace{.12in} \noindent \textbf{3.1. The general algebraic
formalism  for quantum systems}

\vspace{.1in}  Formally, a  \emph{quantum system} is a supmech
Hamiltonian system $(\sca, \sone, \omega, H)$ in which the system
algebra \sca \ is non-supercommutative and $ \omega $ is the
\emph{quantum symplectic form} $ \omega_Q$ given by [see Eq.(44) of
I]
\begin{eqnarray}
\omega_Q = -i \hbar \omega_c
\end{eqnarray}
where $\omega_c$ is the canonical 2-form of \sca \ defined by
Eq.(39) of I (i.e. $\omega_c(D_A,D_B) = [A,B]$). 
[We have, in the terminology of section 3.3 of I, the
quantum symplectic structure with parameter $ b = -i \hbar$. If the
superalgebra \sca \ is not `special' (i.e. not restricted to have
only inner superderivations), we have a generalized symplectic
structure as mentioned at the end of section 4 in I.] This is the
only place where we put the Planck constant `by hand' (the most
natural place to do it --- such a parameter is \emph{needed} here to
give the symplectic form $\omega_Q$ the dimension of action); its
appearance at all conventional places (canonical commutation
relations, Heisenberg and Schr$\ddot{o}$dinger equations, etc) will
be automatic.

The \emph{quantum Poisson bracket} implied by the quantum symplectic
form (35) is [see Eq.(43) of I]
\begin{eqnarray}
\{ A, B \} = (-i\hbar)^{-1}[A, B].
\end{eqnarray}
Recalling that the bracket [,] represents a supercommutator, the
bracket on the right in Eq.(36) is an anticommutator when both A and
B are odd/fermionic and a commutator in all other situations with
homogeneous A,B.

A \emph{quantum canonical transformation} is an automorphism $\Phi$
of the system algebra \sca \ such that $\phstup \omega_Q =
\omega_Q$. Now, by Eq.(12) of I,
\begin{eqnarray}
(\phstup \omega_Q )(X_1, X_2) = \Phi^{-1}[ \omega_Q (\phst X_1,
\phst X_2)]
\end{eqnarray}
where $ X_1, X_2$ are inner superderivations, say, $D_A$ and $D_B$.
We have [recalling Eq.(3) of I]
\begin{eqnarray*}
(\phst D_A)(B) = \Phi [D_A (\Phi^{-1}(B)] = \Phi ( [A,
\Phi^{-1}(B)])
 = [\Phi(A),B] \end{eqnarray*}
which gives
\begin{eqnarray}
\phst D_A = D_{\Phi(A)}.
\end{eqnarray}
Eq.(37) now gives
\begin{eqnarray}
\Phi (i [A,B]) = i [\Phi (A), \Phi (B)]
\end{eqnarray}
which shows, quite plausibly, that quantum canonical transformations
are (in the present algebraic setting --- we have not yet come to
the Hilbert space) the automorphisms of the system algebra
preserving the quantum PBs.

The evolution of a quantum system in time is governed, in the
Heisenberg picture, by the noncommutative Hamilton's equation (49)
of I which now becomes the familiar \emph{Heisenberg equation} of
motion
\begin{eqnarray}
\frac{ dA(t)}{dt} = (-i \hbar)^{-1} [H, A(t)].
\end{eqnarray}
In the Schr$\ddot{o}$dinger  picture, the time dependence is carried
by the states and the evolution equation (51) of I takes the form
\begin{eqnarray}
\frac{d \phi (t)}{dt}(A) = (-i \hbar)^{-1} \phi(t)([H, A])
\end{eqnarray}
which may be called the \emph{generalized von Neumann equation}.

We shall call two quantum systems $\Sigma = (\sca, \sone, \omega,
H)$ and $\Sigma^{\prime} = \linebreak (\sca^{\prime},
\sone^{\prime}, \omega^{\prime}, H^{\prime})$ \emph{equivalent} if
they are equivalent as noncommutative Hamiltonian systems. (See
section 3.4 of I.)

\vspace{.1in} \noindent \emph{Note}. In the abstract algebraic
framework, the CC condition is to be kept track of. We shall see in
the following subsection that this condition permits us to obtain
Hilbert space based realizations of quantum systems (which have the
CC condition built in them as shown in section 2.2 above).

\vspace{.12in} \noindent \textbf{3.2. Inevitability of the Hilbert
space}

\vspace{.12in} Given a quantum system $\Sigma =
(\sca,\sone,\omega,H)$, any other quantum system $\Sigma^{\prime} =
(\sca^{\prime}, \sone^{\prime}, \omega^{\prime}, H^{\prime})$,
equivalent to $\Sigma$ as a noncommutative Hamiltonian system, is
physically equivalent to $\Sigma$ and may be called a realization of
$\Sigma$. By a Hilbert space realization of $\Sigma$ we mean an
equivalent  quantum system $\hat{\Sigma} =
(\hat{\sca},\hat{\sone},\hat{\omega},\hat{H})$ in which $\hat{\sca}$
is an Op*-algebra based on a pair $ (\hat{\sch}, \hat{\mathcal{D}})$
 thus constituting a quantum triple $(\hat{\sch}, \hat{\mathcal{D}},
 \hat{\sca})$ where, in general, the action of $\hat{\sca}$ on
$\hat{\sch}$ need not be irreducible. From the above definition it
is clear that, such a realization, if it exists, is unique up to
equivalence. The precise statement about the existence of these
realizations appears in theorem (1) below.

Construction of the quantum triple $(\hat{\sch}, \hat{\mathcal{D}},
 \hat{\sca})$ is the problem of obtaining a faithful *-representation
of the *-algebra \sca. Some good references for the treatment of
relevant mathematical concepts are (Powers [39], Dubin and Hennings
[20], Horuzhy [28]). By a *-representation of a *-algebra \sca \ we
mean a triple $(\sch, \mathcal{D}, \pi)$ where \sch \ is a
(separable) Hilbert space, $\mathcal{D}$ a dense linear subset of
\sch \ and $\pi$ a *-homomorphism of \sca \ into the operator
algebra $L^{+}(\mathcal{D})$ (the largest *-algebra of operators on
\sch \ having $\mathcal{D}$ as an invariant domain) satisfying the
relation
\begin{eqnarray*} (\chi, \pi(A) \psi) = (\pi(A^*) \chi, \psi) \ \
\textnormal{for all} \ \ A \in \sca \ \textnormal{and} \ \  \chi,
\psi \in \mathcal{D}. \end{eqnarray*} The operators $\pi(A)$ induce
a topology on $\mathcal{D}$ defined by the seminorms
 $ \| . \|_S $ (where S is any finite subset of \sca) given by
\begin{eqnarray}
\|\psi \|_S  =  \sum_{A \in S} \| \pi (A) \psi \|
\end{eqnarray}
where $ \|.\|$ is the Hilbert space norm. The mappings $\pi(A) :
\mathcal{D} \rightarrow \mathcal{D} $ are continuous in this
topology for all $A \in \sca$. The representation $\pi$ is said to
be closed if $\mathcal{D}$ is complete in the induced topology.
Given a *-representation $\pi$ of \sca, there exists a unique
minimal closed extension $\bar{\pi}$ of $\pi$ (called the closure of
$\pi$).

The representation $\pi$ is said to be irreducible if its weak
commutant $\pi_w^{\prime}(\sca)$, defined as the set of bounded
operators C on \sch \ satisfying the condition
\[ (C^* \psi, A \chi) = (A^* \psi, C \chi) \ \ \textnormal{for all}
\  A \in \sca \ \ \textnormal{and} \ \ \psi, \chi \in \mathcal{D} \]
consists of complex multiples of the unit operator.

Once we have the triple $(\hat{\sch}, \hat{\mathcal{D}}, \hat{\pi})$
where $\hat{\pi}$ is a faithful *-representation of \sca, we have
the quantum triple  $(\hat{\sch}, \hat{\mathcal{D}},\hat{\sca})$
where $\hat{\sca} = \hat{\pi}(\sca)$. The construction of $
\hat{\omega}$ and $\hat{H}$ is then immediate :
\begin{eqnarray}\hat{\omega} = -i \hbar \hat{\omega}_c, \ \ \hat{H} =
\hat{\pi}(H) \end{eqnarray} where $\hat{\omega}_c$ is the canonical
form on $\hat{\sca}$. The construction of the Hilbert space-based
realization of the quantum system $\Sigma$ is then completed by
obtaining $\hat{\sone} = \sone (\hat{\sca})$ such that the pair
$(\mathcal{O}(\hat{\sca}), \hat{\sone})$ satisfies the CC condition.

We shall build up our arguments such that no new assumptions will be
involved in going from the abstract algebraic setting to the Hilbert
space setting; emergence of the Hilbert space formalism will be
automatic.

To this end, we shall exploit the fact that the CC condition
guarantees the existence of plenty of (pure) states of the algebra
\sca. Given a state $\phi$ on \sca, a standard way to obtain a
representation of \sca \ is to employ the so-called GNS
construction. Some essential points related to this construction are
recalled below :

\vspace{.1in}  \noindent (i) Considering the given algebra \sca \ as
a complex vector space, one tries to define a scalar product on it
using the state $\phi$ : $(A, B) = \phi(A^*B)$. This, however, is
not positive definite if the set
\begin{eqnarray} L_{\phi} = \{ A \in \sca; \ \phi(A^*A) = 0 \}
\end{eqnarray}
(which can be shown to be a left ideal of \sca) has nonzero elements
in it. On the quotient space $\mathcal{D}^{(0)}_{\phi} =
\sca/L_{\phi}$, the object
\begin{eqnarray} ([A], [B]) = \phi (A^*B) \end{eqnarray}
is a well defined scalar product. Here $ [A] = A + L_{\phi}$ denotes
the equivalence class of A in $\mathcal{D}^{(0)}_{\phi}$. One then
completes the inner product space $(\mathcal{D}^{(0)}_{\phi}, (,))$
to obtain the Hilbert space $\sch_{\phi}$; it is separable if the
topological algebra \sca \ is separable.

\noindent (ii) One obtains a representation $\pi^{(0)}_{\phi}$ of
\sca \ on the pair $(\sch_{\phi}, \mathcal{D}^{(0)}_{\phi})$ by
putting
\begin{eqnarray} \pi^{(0)}_{\phi}(A)[B] = [AB]; \end{eqnarray}
it can be easily checked to be a well defined *-representation. We
denote by $\pi_{\phi}$  the closure of the representation
$\pi^{(0)}_{\phi}$; the completion $\mathcal{D}_{\phi}$ of
$\mathcal{D}^{(0)}_{\phi}$ in the $\pi^{(0)}_{\phi}$-induced
topology acts as the common invariant domain for the operators
$\pi_{\phi}(A)$.

\noindent (iii) The original state $\phi$ is represented as a vector
state in the representations $\pi^{(0)}_{\phi}$ and $\pi_{\phi}$ by
the vector $\chi_{\phi} = [I]$ (the equivalence class of the unit
element of \sca); indeed, we have, from equations (45) and (46),
\begin{eqnarray} \phi(A) & = & ([I], [A]) = ([I],
\pi^{(0)}_{\phi}(A)[I]) \nonumber \\
& = & (\chi_{\phi}, \pi^{(0)}_{\phi}(A) \chi_{\phi}) = (\chi_{\phi},
\pi_{\phi}(A) \chi_{\phi}). \end{eqnarray} The triple $(
\sch_{\phi}, \mathcal{D}_{\phi}, \pi_{\phi})$ satisfying Eq.(47) is
referred to as the GNS representation of \sca \ induced by the state
$\phi$; it is determined uniquely, up to unitary equivalence, by the
state $\phi$. It is irreducible if and only if the state $\phi$ is
pure.

\vspace{.1in} This construction (on a single state), however, does
not completely solve our problem because a GNS representation is
generally not faithful; for all $A \in L_{\phi}$, we have obviously
$\pi_{\phi}(A) = 0$. It is faithful if the state $\phi$ is faithful
(i.e. if $L_{\phi} = \{ 0 \}$). Such a state, however, is not
guaranteed to exist by our postulates.

\vspace{.1in} A faithful but generally reducible representation of
\sca \ can be obtained by taking the direct sum of the
representations of the above sort corresponding to \emph{all} the
pure states $\phi$. [For the construction of the direct sum of a
possibly uncountable set of Hilbert spaces, see (Rudin [41]).] Let
$\mathcal{K}$ be the Cartesian product of the Hilbert spaces $ \{
\sch_{\phi}: \phi \in \sone(\sca) \}$. A general element $\psi$ of
$\mathcal{K}$ is a collection $ \{ \psi_{\phi} \in \sch_{\phi}; \phi
\in \sone(\sca) \}$; here $\psi_{\phi}$ will be called the component
of $\psi$ in $\sch_{\phi}$. The desired Hilbert space \sch \
consists of those elements $\psi$ in $\mathcal{K}$ which have an at
most countable set of nonzero components $ \psi_{\phi}$ which,
moreover, satisfy the condition
\[ \sum_{\phi} \| \psi_{\phi} \|^2_{\sch_{\phi}} < \infty. \]
The scalar product in \sch \ is given by
\[ (\psi, \psi^{\prime}) = \sum_{\phi} (\psi_{\phi},
\psi^{\prime}_{\phi})_{\sch_{\phi}}. \] The direct sum of the
representations $\{ (\sch_{\phi}, \mathcal{D}_{\phi}, \pi_{\phi});
\phi \in \sone(\sca) \} $ is the representation $(\sch, \mathcal{D},
\pi)$ where \sch \ is as above, $\mathcal{D}$ is the subset of \sch
\ consisting of vectors $\psi$ with $ \psi_{\phi} \in
\mathcal{D}_{\phi}$ for all $\phi \in \sone(\sca)$ and, for any $A
\in \sca$,
\[ \pi(A)\psi = \{ \pi_{\phi}(A)\psi_{\phi}; \phi \in \sone(\sca)
\}. \]

Now, given any two different elements $A_1, A_2$ in \oa, let
$\phi_0$ be a pure state (guaranteed to exist by the CC condition)
such that $\phi_0(A_1) \neq \phi_0(A_2)$. Let $ \psi_0 \in \sch$ be
the vector with the single nonzero component $(\psi_0)_{\phi_0} =
\chi_{\phi_0}$. For any $A \in \sca$, we have
\[ (\psi_0, \pi(A) \psi_0) = ( \chi_{\phi_0}, \pi_{\phi_0}(A)
\chi_{\phi_0}) = \phi_0(A). \] This implies
\[ (\psi_0, \pi(A_1) \psi_0) \neq (\psi_0, \pi(A_2) \psi_0), \ \
\textnormal{hence} \ \pi(A_1) \neq \pi(A_2) \] showing that the
representation $(\sch, \mathcal{D}, \pi)$ is faithful.

\vspace{.1in} The Hilbert space \sch \ obtained above may be
non-separable (even if the spaces $\sch_{\phi}$ are separable); this
is because the set $\sone(\sca)$ is generally uncountable. To obtain
a faithful representation of \sca \ on a separable Hilbert space, we
shall use the separability of \sca \ as a topological algebra. Let $
\sca_0 = \{ A_1, A_2, A_3,...\}$ be a countable dense subset of \sca
\ consisting of nonzero elements. The CC condition guarantees the
existence of pure states $\phi_j$ (j=1,2,...) such that
\begin{eqnarray} \phi_j (A_j^* A_j) \neq 0, \ \ j= 1,2,...
\end{eqnarray} Now consider the GNS representations $(\sch_{\phi_j},
\mathcal{D}_{\phi_j}, \pi_{\phi_j})$ (j=1,2,...). Eq.(48) guarantees
that
\begin{eqnarray} \pi_{\phi_j}(A_j) \neq 0 , \ \ j= 1,2,...
\end{eqnarray} Indeed
\begin{eqnarray*} 0 \neq \phi_j(A_j^*A_j) & = & (\chi_{\phi_j},
\pi_{\phi_j}(A_j^*A_j) \chi_{\phi_j}) \\
& = & (\pi_{\phi_j}(A_j) \chi_{\phi_j},
\pi_{\phi_j}(A_j)\chi_{\phi_j}). \end{eqnarray*} Now consider the
direct sum $(\sch^{\prime}, \mathcal{D}^{\prime}, \pi^{\prime})$ of
these representations. To show that $\pi^{\prime}$ is faithful, we
must show that, for any nonzero element A of \sca, $ \pi^{\prime}(A)
\neq 0.$ This is guaranteed by Eq.(49) because, $\sca_0$ being dense
in \sca, A can be arranged to be as close as we like to some $A_j$
in $\sca_0$.

The representation $\pi^{\prime}$, is, in general, reducible. To
obtain a faithful irreducible representation, we should try to
obtain the relations $\pi(A_j) \neq 0$ (j= 1,2,..) in a single GNS
representation $\pi_{\phi}$ for some $\phi \in \sone(\sca)$. To this
end, let $B^{(k)} = A_1 A_2...A_k$ and choose $\phi^{(k)} \in
\sone(\sca)$ such that
\[ \phi^{(k)} (B^{(k)*}B^{(k)}) \neq 0. \]
In the GNS representation $(\sch_{\phi^{(k)}},
\mathcal{D}_{\phi^{(k)}}, \pi_{\phi^{(k)}})$, we have
\[ 0 \neq \pi_{\phi^{(k)}}(B^{(k)}) = \pi_{\phi^{(k)}}(A_1)...
\pi_{\phi^{(k)}}(A_k) \] which implies
\begin{eqnarray} \pi_{\phi^{(k)}}(A_j) \neq 0 , \ \ j=1,...,k.
\end{eqnarray}
This argument works for arbitrarily large but finite k. If the $ k
\rightarrow \infty $ limit of the above construction leading to a
limiting GNS representation $(\underline{\sch},
\underline{\mathcal{D}}, \underline{\pi})$ exists, giving
\begin{eqnarray} \underline{\pi}(A_j) \neq 0, \ \ j= 1,2,...,
\end{eqnarray}
then, by an argument similar to that for $\pi^{\prime}$ above, one
must have $\underline{\pi}(A) \neq 0 $ for all non-zero A in \sca \
showing faithfulness of $\underline{\pi}$.

\vspace{.1in} \noindent \emph{Note.} For system algebras generated
by a finite number of elements (this covers all applications of QM
in atomic physics), a limiting construction is not needed; the
validity of Eq.(50) for sufficiently large k is adequate. [Hint :
Take the generators of the algebra \sca \ as some of the elements of
$\sca_0$.]

 \vspace{.1in} Coming back to the general case, we have, finally,
the faithful (but generally not irreducible) representation
($\hat{\sch}, \hat{\mathcal{D}}, \hat{\pi}$) of \sca \ giving the
desired quantum triple $(\hat{\sch}, \hat{\mathcal{D}}, \hat{\sca})$
where $\hat{\sca} = \hat{\pi}(\sca)$. Since $\hat{\pi}$ is faithful,
$\hat{\sca}$ is an isomorphic copy of \sca. There is a bijective
correspondence $\phi \leftrightarrow \hat{\phi}$ between
$\scs(\sca)$ and $\scs(\hat{\sca})$ [restricting to a bijection
between \sone(\sca) and $\sone(\hat{\sca}) \equiv \hat{\sone}]$ such
that
\begin{eqnarray} <\hat{\phi}, \hat{A}> \ = \ <\phi, A> \ \
\textnormal{for all}\  A \in \sca \end{eqnarray} where $\hat{A} =
\hat{\pi}(A).$ This equation implies that, since the pair
$(\mathcal{O}(\sca), \sone)$ satisfies the CC condition, so will the
pair $(\mathcal{O}(\hat{\sca}), \hat{\sone}).$ We have, finally, a
Hilbert space realization $\hat{\Sigma} =
(\hat{\sca},\hat{\sone},\hat{\omega},\hat{H})$ of the quantum system
$\Sigma = (\sca,\sone,\omega,H).$

Note, from Eq.(52), that
\begin{eqnarray} \hat{\phi} = (\hat{\pi}^{-1})^T(\phi).
\end{eqnarray} When $\hat{\pi}$ is irreducible (equal to
$\pi_{\phi_0}$, say, where $\phi_0 \in \sone(\sca)$), pure states of
$\hat{\sca}$ are vector states $\hat{\phi}_{\psi}$ corresponding to
normalized vectors $\psi \in \hat{\scd}$ :
\begin{eqnarray}
\hat{\phi}_{\psi} (\hat{A}) = (\psi, \hat{A} \psi) = (\psi,
\hat{\pi}(A) \psi). \end{eqnarray} These normalized vectors are of
the form
\begin{eqnarray} \psi_B = N_B^{1/2} [B], \ \ B \in \sca, \ \
B \notin L_{\phi_0}\end{eqnarray} [see equations (45) and (46)]
where $ N_B = [\phi_0(B^* B)]^{-1}.$ Putting $\hat{\phi} =
\hat{\phi}_{\psi_B}$ in Eq.(52), we have
\begin{eqnarray} < \phi,A > & = & < \hat{\phi}_{\psi_B}, \hat{A} > \ =
(\psi_B, \hat{A} \psi_B) = N_B \ ([B], \hat{\pi}(A) [B]) \nonumber \\
& = & N_B \ \phi_0(B^* A B) \equiv \phi_B (A) \end{eqnarray} where
we have defined the linear functional $\phi_B$ on \sca \ by
\begin{eqnarray} \phi_B(A) = N_B \ \phi_0 (B^*AB) \hspace{.15in}
\textnormal{for all} \  A \in \sca. \end{eqnarray} Equations (53)
and (56) now give \begin{eqnarray} \hat{\phi}_{\psi_B} =
(\hat{\pi}^{-1})^T (\phi_B) \ \ \textnormal{for all} \  B \in \sca,
B \notin L_{\phi_0}. \end{eqnarray}

It is instructive to verify directly that the objects $\phi_B(A)$ of
Eq.(57) depend only on the equivalence class [B] and are genuine
elements of \sone(\sca) when $\phi_0 \in \sone(\sca)$.

\vspace{.1in} \noindent \textbf{Proposition 3.1} \emph{Given the
pair (\sca, \sone) of the system algebra \sca \ and its set of pure
states \sone, a state $\phi \in \sone$ and an element $B \in \sca$
such that $ B \notin L_{\phi}$, the linear functional $\phi_B : \sca
\rightarrow \mathbb{C}$ defined by Eq.(57) (with $\phi_0$ replaced
by $\phi$) (a) depends only on the equivalence class $ [B] \equiv B
+ L_{\phi}$ of B,  and (b) is a pure state of \sca.}

\vspace{.1in} \noindent \emph{Proof.} (a) We must show that, for all
$K \in L_{\phi}$ and all $A \in \sca$,
\[ \phi_B(A) = \phi_{B+K}(A) = N_{B+K} \phi [(B+K)^* A (B+K)]. \]
This is easily seen by using the Schwarz inequaliy
\[ |\phi(C^*D)|^2 \leq \phi(C^*C) \ \phi (D^*D) \ \ \textnormal{for
all} \  C, D \in \sca \] and the relation $\phi(K^*K) = 0.$

\noindent (b) Positivity and normalization of the functional
$\phi_B$ are easily proved showing that it is a state. [Note that
the positivity of $\phi_B$ holds only with the convention $ (AB)^* =
B^* A^*$ and not with $(AB)^* = (-1)^{\epsilon_A \epsilon_B} B^*
A^*$; see the note in the beginning of section 2.] To show that it
is a pure state, we shall prove that the GNS representation
$(\sch_B, \mathcal{D}_B,\pi_B)$ induced by the state $\phi_B$ is
unitarily equivalent to the GNS representation $(\sch, \mathcal{D},
\pi)$ induced by the pure state $\phi$ (and is, therefore,
irreducible).

Writing, for $A, B \in \sca,$
\[ [A] \equiv A + L_{\phi}, \ [A]_B \equiv A + L_{\phi_B}, \ \chi =
[I], \ \chi_B = [I]_B ,\] we have
\begin{eqnarray} (\chi_B, \pi_B(A) \chi_B)_{\sch_B} & = & \phi_B (A)
=  N_B \phi (B^* A B)  \nonumber \\ & = & N_B (\chi, \pi (B^* A B)
\chi)_{\sch}. \end{eqnarray}

The object $ \psi_B $ of Eq.(55) is a normalized vector in
$\mathcal{D}.$ Since $ \pi$ is irreducible, the set $\{ \pi (A)
\psi_B; A \in \sca \}$ (with B fixed) is dense in $\mathcal{D}.$
Moreover, the set $ \{ \pi_B(A) \chi_B; A \in \sca \}$ is dense in
$\mathcal{D}_B.$ We define a mapping $U : \mathcal{D} \rightarrow
\mathcal{D}_B$ by
\begin{eqnarray} U \pi(A) \psi_B = \pi_B(A) \chi_B \ \
\textnormal{for all} \ A \in \sca. \end{eqnarray} Now, with $B \in
\sca$ fixed and any $A,C \in \sca$, we have
\begin{eqnarray} (\pi_B(A) \chi_B, \pi_B(C) \chi_B)_{\sch_B} & = &
(\chi_B, \pi_B(A^* C) \chi_B)_{\sch_B} \nonumber \\
& = & N_B (\chi, \pi(B^*A^*CB) \chi)_{\sch} \nonumber \\
& = & (\psi_B, \pi(A^*C) \psi_B)_{\sch} \nonumber \\
& = & (\pi(A) \psi_B, \pi(C) \psi_B)_{\sch} \end{eqnarray} showing
that U is an isometry; by standard arguments, it extends to a
unitary mapping from \sch \ to $\sch_B$ mapping $\mathcal{D}$ onto
$\mathcal{D_B}.$ This proves the desired unitary equivalence of
$\pi$ and $\pi_B$ implying that $\phi_B$ is a pure state. \ \ $\Box$

The proof of part (b) above has yielded a useful corollary :

\vspace{.1in} \noindent \textbf{Corollary (3.2).} \emph{The GNS
representations induced by the states $\phi$ and $\phi_B$ of
proposition (3.1) are related through a unitary mapping as in
Eq.(60).}

\vspace{.1in} Having obtained  the quantum triple $(\hat{\sch},
\hat{\mathcal{D}}, \hat{\sca})$ with the locally convex topology on
$\hat{\mathcal{D}}$ as described above, a mathematically rigorous
version of Dirac's bra-ket formalism (Roberts [40], Antoine [3], A.
B$\ddot{o}$hm [11], de la Madrid [18]) based on the Gelfand triple
\begin{eqnarray} \hat{\mathcal{D}} \subset \hat{\sch} \subset
\hat{\mathcal{D}}^{\prime} \end{eqnarray} where
$\hat{\mathcal{D}}^{\prime}$ is the dual space of
$\hat{\mathcal{D}}$ with the strong topology (Kristensen, Mejlbo and
Thue Poulsen [30]) defined by the seminorms
 $p_W$ given by \[ p_W (F) = sup_{\psi \in W} \ | F(\psi)| \ \
 \textnormal{for all} \ F \in \hat{\mathcal{D}}^{\prime} \] for all bounded
 sets W of $\hat{\mathcal{D}}$; the triple (62) constitutes the
 \emph{canonical rigged Hilbert space} based on $(\hat{\sch},
\hat{\mathcal{D}})$ (Lassner [31]). The space
$\hat{\mathcal{D}}^{\prime}$ ( the space of continuous linear
functionals on  $\hat{\mathcal{D}}$) is the space of bra vectors of
Dirac. The space of kets is the space $\hat{\mathcal{D}}^{\times}$
of continuous antilinear functionals on $\hat{\mathcal{D}}$. [An
element $\chi \in \sch$ defines a continuous linear functional
$F_{\chi}$ and an antilinear functional $K_{\chi}$ on $\hat{\sch}$ \
(hence on $\hat{\mathcal{D}}$) given by $ F_{\chi}(\psi) = (\chi,
\psi)$ and $K_{\chi}(\psi) = (\psi, \chi)$; both the bra and ket
spaces, therefore, have \sch \ as a subset.]

When $\hat{\pi}$ is irreducible, the (unnormalized) vectors in
$\hat{\mathcal{D}}$ representing pure states of $\hat{\sca}$ have
unrestricted superpositions allowed between them; they constitute a
coherent set in the sense of (Bogolubov [10]) (which means that
they, as a set,  cannot be represented as a union of two nonempty
mutually orthogonal sets). We can now follow the reasoning employed
in the proof of lemma (4.2) in (Bogolubov [10]) to conclude that, in
the general case (when $\hat{\pi}$ may be reducible), the Hilbert
space $\hat{\sch}$ can be expressed as a direct sum of mutually
orthogonal coherent subspaces :
\begin{eqnarray} \hat{\sch} = \bigoplus_{\alpha} \hat{\sch}_{\alpha}
\end{eqnarray} such that each of the  $\hat{\mathcal{D}}_{\alpha} \equiv
\hat{\mathcal{D}} \cap \sch_{\alpha} $ is a coherent set on which
$\hat{\sca}$ acts irreducibly (but not necessarily faithfully) and
$\hat{\mathcal{D}} = \cup_{\alpha} \hat{\mathcal{D}}_{\alpha}$.
[Introduce an equivalence relation  $\sim$ in $\hat{\mathcal{D}}$ :
$\psi \sim \chi$ if there is a coherent subset $\mathcal{C}$ in
$\hat{\mathcal{D}}$ to which both $\psi, \chi$ belong. This gives
the equivalence classes $\hat{\mathcal{D}}_{\alpha}$ in
$\hat{\mathcal{D}}$. Define $ \hat{\sch}_{\alpha}$ as the closure of
$\hat{\mathcal{D}}_{\alpha}$ in $\hat{\sch}$, etc.] The breakup (63)
implies the breakup $ \hat{\pi} = \oplus_{\alpha}
\hat{\pi}_{\alpha}$ where each triple $(\hat{\sch}_{\alpha},
\hat{\mathcal{D}}_{\alpha}, \hat{\pi}_{\alpha})$ is an irreducible
(but not necessarily faithful) representation of \sca. For every $A
\in \sca$ and $\psi = \{ \psi_{\alpha} \in
\hat{\mathcal{D}}_{\alpha} \} \in \hat{\mathcal{D}}$, we have
\begin{eqnarray} \hat{\pi}(A) \psi = \{ \hat{\pi}_{\alpha}(A)
\psi_{\alpha} \}. \end{eqnarray} This situation corresponds to the
existence of superselection rules; the subspaces
$\hat{\sch}_{\alpha}$ are referred to as coherent subspaces or
superselection sectors. The projection operators $P_{\alpha}$ for
the subspaces $\hat{\sch}_{\alpha}$ belong to the center of
$\hat{\sca}$. [To show this, it is adequate to show that, for any
$\hat{A} \equiv \hat{\pi}(A) \in \hat{\sca}$ and $\psi =
\{\psi_{\alpha} \} \in \hat{\mathcal{D}}, \ \hat{A}P_{\alpha} \psi =
P_{\alpha} \hat{A} \psi. $ Using Eq.(64), each side is easily seen
to be equal to $\hat{\pi}_{\alpha} (A) \psi_{\alpha}$.]

Operators of the form
\begin{eqnarray} Q = \sum_{\alpha} a_{\alpha} P_{\alpha}, \ \
a_{\alpha} \in \mathbb{R} \end{eqnarray} serve as superselection
operators. Any two such operators obviously commute. We have,
therefore, a formalism in which there is a natural place for
superselection rules which are restricted to be commutative.

We have proved the following theorem.

\vspace{.1in} \noindent \textbf{Theorem(1).} \emph{Given a quantum
system $\Sigma = (\sca,\sone,\omega,H)$ (where the system algebra
\sca \ is supposedly  separable as a topological algebra), the
following holds true.}

\noindent (a) \emph{The system algebra \sca \ admits a faithful
*-representation $(\hat{\sch},
\hat{\mathcal{D}}, \hat{\pi})$ in a separable Hilbert space
$\hat{\sch}$ giving  the quantum triple $(\hat{\sch},
\hat{\mathcal{D}}, \hat{\sca})$ with $\hat{\sca}= \hat{\pi}(\sca)$.}

\noindent (b) \emph{With pure states defined through Eq.(53) and the
quantum symplectic form $\hat{\omega}$ and the Hamiltonian operator
$\hat{H}$ given by Eq.(43), this provides the Hilbert space based
realization $\hat{\Sigma} =
(\hat{\sca},\hat{\sone},\hat{\omega},\hat{H})$ of the quantum system
$\Sigma$. This realization supports a rigorous version of the Dirac
bra-ket formalism based on the canonical rigged Hilbert space (62).}

\noindent (c) \emph{When \sca \ is generated by a finite number of
elements, it is possible to have the faithful *-repesentation
$\hat{\pi}$ of part (a) irreducible. In this case pure states of
$\hat{\mathcal{A}}$ are the vector states corresponding to the
normalized elements of $\hat{\mathcal{D}}$.}

\noindent (d) \emph{In the general case, the Hilbert space
$\hat{\sch}$  of (a) above can be expressed as a direct sum (63) of
mutually orthogonal subspaces (superselection sectors) such that
each $\sch_{\alpha}$ is an irreducible invariant subspace for the
opertor algebra $\hat{\sca}$,  each set $\mathcal{D}_{\alpha}$ is
coherent and $\hat{\mathcal{D}} = \cup_{\alpha}
\mathcal{D}_{\alpha}$. The superselection operators (65) constitute
a real subalgebra of the center of $\hat{\sca}$.}

\vspace{.1in} We shall call a quantum system with a finitely
generated system algebra a \emph{standard quantum system}. According
to theorem (1), such a system admits a Hilbert space based
realization with the system algebra represented faithfully and
irreducibly and there are no superselection rules. All quantum
systems consisting of a finite number of particles (in particular
all quantum systems in atomic physics) obviously belong to this
class.

\vspace{.12in} \noindent \textbf{3.3. Hilbert space quantum
mechanics of standard quantum systems}

\vspace{.11in} We shall now consider Hilbert space based
realizations of standard quantum systems and relate the supmech
treatment of their kinematics and dynamics in section 3.1 to the
traditional Hilbert space based formalism.

We first consider the implementation of symplectic mappings in such
realizations. The main result is contained in the following theorem.

\vspace{.12in} \noindent \textbf{Theorem (2).} \emph{Let $\Sigma =
(\sca,\sone,\omega,H)$ and $\Sigma^{\prime} =
(\sca^{\prime},\sone^{\prime},\omega^{\prime},H^{\prime})$ be two
equivalent standard quantum systems; the equivalence is described by
the symplectic mappings $\Phi = (\Phi_1, \Phi_2)$ [which means that
$\Phi_1 : \sca \rightarrow \sca^{\prime}$ is an isomorphism of
unital *-algebras  such that $\Phi^* \omega^{\prime} = \omega$ and
$\Phi_2 : \sone \rightarrow \sone^{\prime}$ is a bijection such that
$<\Phi_2(\phi), \Phi_1(A)> \ = \ <\phi,A>$ for all $\phi \in \sone$
\ and $A \in \sca$]. Given their Hilbert space realizations
$\hat{\Sigma} = (\hat{\sca},\hat{\sone},\hat{\omega},\hat{H})$ and
$\hat{\Sigma}^{\prime} =
(\hat{\sca}^{\prime},\hat{\sone}^{\prime},\hat{\omega}^{\prime},
\hat{H}^{\prime})$ [the respective representations of system
algebras being ($\hat{\sch}, \hat{\mathcal{D}}, \hat{\pi}$) and
($\hat{\sch}^{\prime}, \hat{\mathcal{D}}^{\prime},
\hat{\pi}^{\prime}$)], there exists a unitary mapping $U :
\hat{\sch} \rightarrow \hat{\sch}^{\prime}$ mapping
$\hat{\mathcal{D}}$ onto $\hat{\mathcal{D}}^{\prime}$ implementing
the given equivalence with} \begin{eqnarray} \hat{\pi}^{\prime}
(\Phi_1(A)) = U \hat{\pi}(A) U^{-1} \ \textnormal{for all} \ A \in
\sca; \ \psi^{\prime} = U \psi
\end{eqnarray} \emph{where $\psi \in \hat{\mathcal{D}}$ and
$\psi^{\prime} \in \hat{\mathcal{D}}^{\prime}$ are representative
vectors for the states $\phi \in \sone$ and $\Phi_2(\phi) \in
\sone^{\prime}$ respectively.}

\vspace{.12in} \noindent \emph{Proof.} Since the quantum systems are
standard, their pure states are represented by normalized vectors in
$\hat{\mathcal{D}}$ and $\hat{\mathcal{D}}^{\prime}$. Let $\phi \in
\sone, \ \phi^{\prime} = \Phi_2(\phi)$  and $\psi \in
\hat{\mathcal{D}}$ and $\psi^{\prime} \in
\hat{\mathcal{D}}^{\prime}$ are normalized vectors such that $
\phi_{\psi} = (\hat{\pi}^{-1})^T(\phi)$ and $ \phi_{\psi^{\prime}} =
([\hat{\pi}^{\prime}]^{-1})^T(\phi^{\prime})$ are the corresponding
vector states in $\hat{\sone}$ and $\hat{\sone}^{\prime}$
respectively. Writing $\hat{A} = \hat{\pi}(A)$  for $A \in \sca$ and
$\hat{A}^{\prime} = \hat{\pi}^{\prime}(A^{\prime})$ for $A^{\prime}
= \Phi_1(A) \in \sca^{\prime}$, we have
\begin{eqnarray} (\psi^{\prime}, \hat{A}^{\prime} \psi^{\prime})_{\hat{\sch}^{\prime}} & =
& <\phi_{\psi^{\prime}}, \hat{\pi}^{\prime}(A^{\prime})> \  =  \ <
\phi, A>  \nonumber \\ & = & <\phi_{\psi}, \hat{\pi}(A)> \ = \
(\psi, \hat{A} \psi)_{\hat{\sch}}
\end{eqnarray}
for all $A \in \sca$ and all $\phi \in \sone$.

Let $\{ \chi_r \}$ (r = 1,2,...) be an orthonormal basis in
$\hat{\sch}$ (with all $\chi_r \in \hat{\mathcal{D}}$), $\phi_r \in
\sone$ \ the state reprented by the vector $\chi_r$,
$\phi_r^{\prime} = \Phi_2(\phi_r)$ and $\chi_r^{\prime} \in
\hat{\mathcal{D}}^{\prime}$ a normalized vector representing the
state $\phi_r^{\prime}$. Define a mapping $U : \hat{\sch}
\rightarrow \hat{\sch}^{\prime}$ such that $U\chi_r =
\chi_r^{\prime}$ (r= 1,2,...). Putting $\psi = \chi_s$ and
$\psi^{\prime} = \chi_s^{\prime}$ in Eq.(67), we have (dropping the
subscripts on the scalar products)
\[ (U \chi_s, \hat{A}^{\prime} U \chi_s) = (\chi_s, \hat{A} \chi_s).
\]
Writing similar equations with $\chi_s$ replaced by $(\chi_r +
\chi_s)/\sqrt 2$ and $(\chi_r + i \chi_s)/\sqrt 2$ we obtain the
relation
\[ (\chi_r, U^{\dagger} \hat{A}^{\prime} U \chi_s) =
(\chi_r, \hat{A} \chi_s) \]
(for arbitrary r and s) which implies
\[  U^{\dagger} \hat{A}^{\prime} U  = \hat{A}  \ \
\textnormal{for all} \  A \in \sca.\] Now, for A = I, we must have $
A^{\prime} = I$ ( the mapping $\Phi_1$ being an isomorphism of the
unital algebra \sca \ onto $\sca^{\prime}$); this gives $U^{\dagger}
U = I $ or, remembering the invertibility of the mapping $\Phi_2$,
$U^{\dagger} = U^{-1}$. We have, therefore, $\hat{A}^{\prime} =
U\hat{A}U^{-1}.$ The condition (39) implies \[ U (i
[\hat{A},\hat{B}]) U^{-1} = i [U\hat{A}U^{-1}, U\hat{B}U^{-1}] \]
which permits U to be taken as a linear and, therefore, unitary
operator.

Now let $\psi = \sum a_r \chi_r$. We have
\[ U \psi = \sum a_r U \chi_r = \sum a_r \chi_r^{\prime} \equiv
\psi^{\prime \prime}.\] This gives, employing the Dirac notation for
projectors,
\[|\psi^{\prime \prime}><\psi^{\prime \prime}| = U
|\psi><\psi|U^{-1} = |\psi^{\prime}><\psi^{\prime}| \] where the
last step follows from Eq.(67) (with $\hat{A}^{\prime} =
U\hat{A}U^{-1}.$) and the CC condition. It follows that
$\psi^{\prime \prime}$ is an acceptable representative of the state
represented by $\psi^{\prime}$ implying that we can consistently
take $\psi^{\prime} = U \psi.$ \ \ $\Box$

\vspace{.1in} We shall say, in the context of the above theorem,
that the mappings $(\Phi_1, \Phi_2)$ are \emph{unitarily
implemented}. Taking $\Sigma^{\prime} = \Sigma$ in the theorem, we
have

\vspace{.12in} \noindent \textbf{Corollary (3.3).} \emph{Given two
Hilbert space realizations $\hat{\Sigma}$ and $\hat{\Sigma}^{\prime}$
of a standard quantum system $\Sigma$, the
mappings describing their equivalence as supmech Hamiltonian systems
can be implemented unitarily.}

\vspace{.12in} Taking $\hat{\Sigma}^{\prime} = \hat{\Sigma}$ in
corollary (3.3), we have

\vspace{.12in} \noindent \textbf{Corollary (3.4).} \emph{In a
Hilbert space realization of a standard quantum system, a quantum
canonical transformation can be implemented unitarily.}

\vspace{.1in} We shall henceforth drop the tildes and take $\Sigma =
(\sca, \sone, \omega, H)$ directly as a Hilbert space realization of
a standard quantum system; here \sca \ is now an Op$^*$-algebra
based on the pair $(\sch, \mathcal{D})$ constituting a quantum
triple $(\sch, \mathcal{D}, \sca)$. In concrete applications, there
is some freedom in the choice of $\mathcal{D}$. When \sca \ is
generated by a finite set of fundamental observables $ F_1,..,F_n,$
 a good choice is, in the notation of Dubin and Hennings [20], $
\mathcal{D} = C^{\infty}(F_1,..,F_n)$ (i.e. intersection of the
domains of all polynomials in $F_1,..,F_n$).

We have now  \sca \ as our system algebra; its states are given by
the subclass of density operators $\rho$ on \sch \ for which $ |Tr
(\bar{\rho A})| < \infty$ (where the overbar indicates closure of
the operator) for all observables A in \sca \ [20]; the quantity
$Tr(\bar{\rho A}) \equiv \phi_{\rho}(A)$ (where $\phi_{\rho}$ is the
state represented by the density operator $\rho$) is the expectation
value of the observable A in the state $\phi_{\rho}$. Pure states
are the subclass of these states consisting of one-dimensional
projection operators $|\psi><\psi|$ where $\psi$ is any normalized
element of $\mathcal{D}$.

The density operators representing states, being Hermitian
operators, are also observables. A density operator $\rho$ is the
observable corresponding to the property of  the system being in the
state $\phi_{\rho}$. Given two states represented by density
operators $\rho_1$ and $\rho_2$, we have the quantity $ w_{12} =
Tr(\rho_1 \rho_2)$ defined (representing the expectation value of
the observable $\rho_1$ in the state $\rho_2$ and  vice versa) which
has the natural interpretation of transition probability  from one
of the states to the other (the two are equal because $ w_{12} =
w_{21}).$ When $ \rho_i = |\psi_i><\psi_i|$ (i = 1,2) are pure
states, we have $Tr(\rho_1 \rho_2) = |(\psi_1,\psi_2)|^2$ --- the
familiar text book expression for the transition probability between
two pure quantum states.

\vspace{.1in} \noindent \emph{Note.} Recalling the stipulation in
section 2.1 about probabilities in the formalism, it is desirable to
represent the quantities $w_{12}$ as bonafide probabilities in the
standard form (1) employing an appropriate PObVM [which, in the
present Hilbert space setting, should be a traditional POVM
(positive operator valued measure)]. It is clearly adequate to have
such a representation for the case of pure states with $\rho_j =
|\psi_j><\psi_j|$ (j = 1,2), say. To achieve this, let $ \phi =
\phi_{\rho_1}$ and $ \{ \chi_r; r = 1,2,...\}$ an orthonormal basis
in \sch \ having $\chi_1 = \psi_2$. The desired POVM is obtained by
taking, in the notation of section 2.1, \begin{eqnarray} \Omega = \{
\chi_r; r = 1,2,...\}, \ \ \mathcal{F} = \{ \textnormal{All subsets
of}\ \Omega \}
\end{eqnarray} and, for $ E = \{ \chi_r; r \in J \} \in \mathcal{F}$
where J is a subset of the positive integers,
\begin{eqnarray} \nu (E) = \sum_{r \in J} |\chi_r><\chi_r|.
\end{eqnarray} We now have
$w_{12} = |(\psi_1, \psi_2)|^2 = p_{\phi}(E)$ of Eq.(1) with $\phi =
\phi_{\rho_1}$ and $E = |\chi_1><\chi_1| = |\psi_2><\psi_2|$.

 \vspace{.1in} The unitarily implemented $\Phi_2$ actions (quantum
canonical transformations) on states leave the transition
probabilities invariant [in fact, they leave transition amplitudes
invariant : $(\psi^{\prime}, \chi^{\prime}) = (\psi, \chi)$]. Note
that, in contrast with the traditional formalism of QM, invariance
of transition probabilities under the fundamental symmetry
operations of the theory is not postulated but proved in the present
setting. The fundamental symmetry operations themselves came as a
matter of course from the basic premises of the theory :
noncommutative symplectics --- exactly as the classical canonical
transformations arise naturally in the traditional commutative
symplectics.

A symmetry implemented (in the unimodal sense, as defined in section
3.4 of I) by a unitary operator U acts on a state vector $\psi \in
\mathcal{D}$ according to $ \psi \rightarrow \psi^{\prime} = U \psi$
and (when its action is transferred to operators) on an operator $A
\in \sca$ according to $A \rightarrow A^{\prime}$ such that, for all
$\psi \in \mathcal{D}$,
\begin{eqnarray}
(\psi^{\prime}, A \psi^{\prime}) = (\psi, A^{\prime} \psi)
\hspace{.15in} \Rightarrow  A^{\prime} = U^{-1} A U .
\end{eqnarray}
For an infinitesimal unitary transformation, $ U \simeq I + i
\epsilon G$ where G is an even, Hermitian element of \sca \ [this
follows from the condition $ (U \phi, U \psi) = (\phi, \psi)$ for
all $\phi, \psi \in \mathcal{D}$]. Considering the transformation $
A \rightarrow A^{\prime}$ in Eq.(70) as a quantum canonical
transformation, generated (through PBs) by an element $ T \in \sca$,
we have
\begin{eqnarray}
\delta A = - i \epsilon [G, A] = \epsilon \{ T, A \}
\end{eqnarray}
giving $ T = -i (-i\hbar) G = - \hbar G $  and
\begin{eqnarray}
U \simeq I -i \frac{\epsilon}{\hbar}T.
\end{eqnarray}
It is the appearance of $\hbar$ in Eq.(72) which is responsible for
its appearance at almost all conventional places in QM.

The quantum canonical transformation representing evolution of the
system in time  is implemented on the state vectors by a
one-parameter family of unitary operators  [in the form $ \psi(t) =
U(t-s)\psi(s)$] generated by the Hamiltonian operator H :
$U(\epsilon) \simeq I -i\frac{\epsilon}{\hbar} H.$ This gives, in
the Schr$\ddot{o}$dinger picture, the Schr$\ddot{o}$dinger equation
for the evolution of pure states :
\begin{eqnarray}
i \hbar \frac{d \psi(t)}{dt} = H \psi(t).
\end{eqnarray}
In the Heisenberg picture, we have, of course, the Heisenberg
equation of motion (40), which is now an operator equation on the
dense domain $\mathcal{D}$.

We had seen in the previous subsection that quantum triples provide
a natural setting for a mathematically rigorous development of the
Dirac bra-ket formalism. For later use, we recall a few points
relating to this formalism which hold good when the space
$\mathcal{D}$ is nuclear (Gelfand and Vilenkin [22]).

A self-adjoint  operator A  in \sca \  in a rigged Hilbert space
(with nuclear rigging as mentioned above) has complete sets of
generalized eigenvectors [eigenkets $\{ |\lambda>; \lambda \in
\sigma (A)$, the spectrum of A $ \}$ and eigenbras $ \{ <\lambda |;
\lambda \in \sigma(A) \}]$ : \begin{eqnarray}  A |\lambda> = \lambda
| \lambda>; \  <\lambda | A = \lambda < \lambda |; \
\int_{\sigma(A)} d \mu(\lambda) |\lambda>< \lambda| = I
\end{eqnarray} where I is the unit operator in \sch \ and $\mu $ is
a unique measure on $\sigma(A)$. These equations are to be
understood in the sense that, for all $\chi, \psi \in \mathcal{D}$,
\[ <\chi| A | \lambda> = \lambda <\chi| \lambda>; \hspace{.2in}
<\lambda | A | \psi> = \lambda <\lambda| \psi>; \] \[
\int_{\sigma(A)} d \mu (\lambda)< \chi| \lambda><\lambda| \psi>\  =
\ <\chi| \psi>. \] The last equation implies the expansion (in
eigenkets of A)
\begin{eqnarray}
| \psi> = \int_{\sigma(A)} d \mu (\lambda) \  |\lambda>< \lambda|
\psi>.
\end{eqnarray}
More generally, one has complete sets of generalized eigenvectors
associated with finite sets of commuting self-adjoint operators.

\vspace{.12in} \noindent \textbf{3.4. Quantum mechanics of
localizable elementary systems (massive particles)}

\vspace{.12in} A \emph{quantum elementary system} is a standard
quantum system which is also an elementary system. The concept of a
quantum elementary system, therefore, combines the concept of
quantum symplectic structure with that of a relativity scheme. The
basic entities relating to an elementary system are its fundamental
observables which generate the system algebra \sca. For quantum
elementary systems, this algebra \sca \ has the quantum symplectic
structure as described in section 3.1. All the developments in
section 2.5 can now proceed with the Poisson brackets (PBs)
understood as quantum PBs of Eq.(36). Since the system algebra is
finitely generated, theorem (1) guarantees the existence of a
Hilbert space-based realization of such a system involving a quantum
triple $(\hat{\sch}, \hat{\mathcal{D}}, \hat{\sca})$ where
$\hat{\sca}$ is a faithful irreducible representation of \sca \
based on the pair $(\hat{\sch}, \hat{\mathcal{D}})$. We shall drop
the hats and call the quantum triple $(\sch, \mathcal{D}, \sca)$.

The  relativity group $G_0$ (or its projective group $\hat{G}_0$)
has a Poisson action on \sca \ and a transitive action on the set
$\sone(\sca)$ of pure states of \sca. We have seen above that, in
the present setting, a symmetry operation can be represented as a
unitary operator on \sch \ mapping $\mathcal{D}$ onto itself. A
symmetry group is then realized as a (projective) unitary representation
on \sch \ having $\mathcal{D}$ as an invariant domain. For an elementary
system the condition of transitive action on \sone \ implies that
this representation must be irreducible. (There is no contradiction
between this requirement and that of invariance of $\mathcal{D}$
because $\mathcal{D}$ is not a closed subspace of \sch \ when \sch \
is infinite dimensional.)

\vspace{.1in} \noindent \emph{Note.} We  now have a formal
justification for the direct route to the Hilbert space taken in the
traditional treatment of QM of elementary systems, namely,
employment of projective unitary irreducible representations of the
relativity group $G_0$. This is the simplest way to simultaneously satisfy the
condition of transitive action of $ G_0$ on the space of pure states
and  the CC condition.

\vspace{.1in} By a (quantum) \emph{particle} we shall mean a
localizable (quantum) elementary system. We shall  consider only
nonrelativistic particles. The configuration space of a
nonrelativistic particle is the 3-dimensional Euclidean space
$\mathbb{R}^3$. The fundamental observables for such a system were
identified, in section 2.5, as the mass (m) and Cartesian components
of  position ($X_j$), momentum ($P_j$) and spin($S_j$) (j = 1,2,3)
satisfying the PB relations in equations (26,27,13). The mass m will
be treated, as before, as a positive parameter. The system algebra
\sca \ of the particle is the
*-algebra generated by the fundamental observables (taken as
hermitian) and the unit element. Since it is an ordinary
*-algebra (i.e. one not having any fermionic objects), the
supercommutators reduce to ordinary commutators. Recalling Eq.(36),
the PBs mentioned above now take the form of the commutation
relations
\begin{eqnarray}
  [X_j, X_k] = 0 = [P_j, P_k], \ \  [X_j, P_k] =
i \hbar \delta_{jk} I \hspace{1.2in} (A)\nonumber \\
 {[S_j, S_k]} = i \hbar \epsilon_{jkl} S_l, \ \ [S_j, X_k] = 0 =
[S_j, P_k]. \hspace{1.4in} (B)
\end{eqnarray}

We now consider explicit construction of the quantum triple $(\sch,
\mathcal{D}, \sca)$ for these objects. We shall first consider the
spinless particles ($\mathbf{S} = 0$); for these, we need to
consider only the Heisenberg commutation relations (76A) [often
referred to as the \emph{canonical commutation relations} (CCR)].
Since the final construction is guaranteed to be unique upto unitary
equivalence, we can allow ourselves to be guided by considerations
of simplicity and plausibility.

Eq.(12), written (with n = 3) for a pure state  (represented by a
normalized vector $\psi \in \mathcal{D}$) now takes the form
(writing $\mu_{\psi}$ for $\mu_{\phi_{\psi}}$)
\begin{eqnarray*} (\psi, X_j \psi) = \int_{\mathbb{R}^3} x_j d
\mu_{\psi} (x) \end{eqnarray*} which shows that the scalar product
in \sch \ involves integration over $\mathbb{R}^3$  with respect to
a measure. The group of space translations is to be represented
unitarily in \sch \ (being a subgroup of the Galilean group). The
simplest choice (which eventually works well as we shall see) is to
take $\sch = L^2(\mathbb{R}^3, dx)$ and the unitary operators U(a)
representing space translations as given by
 \begin{eqnarray}
[U(a) \psi](x) = \psi (x-a)
\end{eqnarray}
[which is a special case of of the relation [$U(g) \psi](x) =
\psi(T_{g}^{-1}x)$; these operators are unitary when the
transformation $T_g$ of $\mathbb{R}^3$ preserves the Lebesgue
measure]. Recalling Eq.(72), we have, for an infinitesimal
translation, $\delta \psi = - \frac{i}{\hbar} \mathbf{a.P} \psi = -
\mathbf{a.\bigtriangledown} \psi $ giving the  operators $P_j$
representing momentum components as
\begin{eqnarray}
(P_j \psi)(x) = -i \hbar \frac{\partial \psi}{\partial x_j}.
\end{eqnarray}
Taking the position operators $X_j$ to be the multiplication
operators given by
\begin{eqnarray}
(X_j \psi)(x) = x_j \psi(x),
\end{eqnarray}
the CCR of Eq.(76A) are satisfied.

We now have [20]
\[ \mathcal{D} = C^{\infty}(X_j, P_j,; j = 1,2,3)  = \scs
(\mathbb{R}^3). \] The operators U(a) clearly map the domain
$\mathcal{D} = \scs (\mathbb{R}^3)$ onto itself. With this choice of
$\mathcal{D}$, the operators $X_j$ and $P_j$ given by equations (79)
and (78) are essentially  self adjoint; we denote their self adjoint
extensions by the same symbols.

The space $\scs(\mathbb{R}^3)$ is nuclear [10] and the rigged
Hilbert space \[ \scs(\mathbb{R}^3) \subset L^2(\mathbb{R}^3)
\subset \scs^{\prime}(\mathbb{R}^3) \] satisfies the conditions for
the validity of the results stated at the end of section 3.3. We
shall make use of the complete sets of generalized eigenvectors of
the operators $X_j$. Let $ x = (x_1, x_2, x_3),\ dx = dx_1 dx_2 dx_3
$ and $|x>, <x|$ the simultaneous eigenkets and eigenbras of the
operators $X_j$ (j= 1,2,3):
\begin{eqnarray}
X_j |x> \ = \ x_j |x>, \ \ <x|X_j \ = \ <x| x_j, \ \ x_j \in
\mathbb{R}, \ \ j=1,2,3;
\end{eqnarray}
they form a complete set providing a resolution of identity in the
form
\begin{eqnarray}
I =  \int_{\mathbb{R}^3} |x> dx <x|.
\end{eqnarray}
Given any vector $ |\psi>  \in \mathcal{D}$, the corresponding wave
function appearing in Eq.(79) is $ \psi(x) \equiv \  <x|\psi>$; we
have, indeed,
\begin{eqnarray*}
(X_j \psi)(x) = \ <x|X_j|\psi> \ = x_j \psi(x).
\end{eqnarray*}

Recalling the discussion of localization in section 2.4, the
localization observable  P(D) corresponding to a Borel set D in
$\mathbb{R}^3$ is represented as  the operator
 \begin{eqnarray}
 P(D) = \int_D |x> dx <x|.
 \end{eqnarray}
[The required properties of P(D) are easily verified.] Given the
particle in the state corresponding to $|\psi> \ \in \mathcal{D}$,
the probability that it will be found in the domain D is given by
\begin{eqnarray}
<\psi| P(D) |\psi> = \int_D <\psi|x> dx <x| \psi> = \int_D
|\psi(x)|^2 dx
\end{eqnarray}
giving the traditional Born interpretation of the wave function
$\psi$.

 The pair  $ (\sch, \mathcal{D}) = (L^2(R^3), \scs(R^3))$ with
operators $X_j$ and $P_j$ as constructed above is known as the
\emph{Schr$\ddot{o}$dinger representation} of the CCR (76A).

 The self adjoint operators $P_j, X_j$ generate the unitary groups
of operators $ U(a) = exp(-ia.P)$ and $V(b) = exp(-ib.X)$ (where
$a.P = \sum_j a_jP_j$ etc. and we have put $\hbar = 1.$) which
satisfy the Weyl commutation relations
\begin{eqnarray}
U(a)U(b) &=& U(b)U(a) = U(a+b), \ V(a)V(b) = V(b)V(a) = V(a+b) \nonumber \\
U(a) V(b) &=& e^{ia.b} V(b) U(a).
\end{eqnarray}
For all $\psi \in \mathcal{D}$, we have
\begin{eqnarray}
(U(a)\psi)(x) = \psi(x-a), \ \ (V(b)\psi)(x) = e^{-ib.x} \psi(x);
\end{eqnarray}
this is referred to as the Schr$\ddot{o}$dinger representation of
the Weyl commutation relations. According to the uniqueness theorem
of von Neumann [44], the irreducible representation of the Weyl
commutation relations is, up to unitary equivalence, uniquely given
by the Schr$\ddot{o}$dinger representation (85).

\vspace{.1in} \noindent \emph{Note}. (i) Not every representation of
the CCR (76A) with essentially self adjoint $X_j$ and $P_j$ gives a
representation of the Weyl commutation relation. [For a
counterexample, see Inoue [29], example (4.3.3).] A necessary and
sufficient condition for the latter to materialize is that the
harmonic oscillator Hamiltonian operator $ H = P^2/(2m) + kX^2/2$ be
essentially self adjoint. In the Schr$\ddot{o}$dinger representation
of the CCR obtained above, this condition is satisfied [23,20]

\vspace{.1in} \noindent (ii) The von Neumann uniqueness theorem
serves to confirm/verify, in the present case, the uniqueness (up to
equivalence) of the Hilbert space realization of a standard quantum
system  mentioned in sections 3.2 and 3.3. Taking the opposite view,
given the uniqueness (up to unitary equivalence) of the Hilbert
space realizations of the algebraic quantum system corresponding to
a nonrelativistic massive spinless particle and the remark (i)
above, we have an alternative proof of the von Neumann uniqueness
theorem.

\vspace{.1in} Quantum dynamics of a free nonrelativistic spinless
particle is governed, in the Schr$\ddot{o}$dinger picture, by the
Schr$\ddot{o}$dinger equation (73) with $\psi \in \mathcal{D} =
\mathcal{S}(\mathbb{R}^3)$ and  with the Hamiltonian (29) [where
$\mathbf{P}$ is now the operator in Eq.(78)]:
\begin{eqnarray} i \hbar \frac{\partial \psi}{\partial t} = -
\frac{\hbar^2}{2m} \bigtriangledown^2 \psi. \end{eqnarray} Explicit
construction of the projective unitary representation of the
Galilean group $G_0$ in the Hilbert space $ \sch =
L^2(\mathbb{R}^3,dx)$ and Galilean covariance of the free particle
Schr$\ddot{o}$dinger equation (86) have been treated in the
literature [5, 44, 16].

When external forces are acting, the Hamiltonian operator has the
more general form (30). Restricting V in this equation to a function
of $\mathbf{X}$ only (as is the case in common applications), and
proceeding as above, we obtain the traditional Schr$\ddot{o}$dinger
equation
\begin{eqnarray}
i \hbar \frac{\partial \psi}{\partial t} = [- \frac{\hbar^2}{2m}
 \bigtriangledown^2 + V(\mathbf{X})] \psi
 \end{eqnarray}
where \textbf{X} is now the position operator of Eq.(79).

\vspace{.1in} It should be noted that, in the process of obtaining
the Schr$\ddot{o}$dinger equation (87) for a nonrelativistic
spinless particle with the traditional Hamiltonian operator, we did
not use the classical Hamiltonian or Lagrangian for the particle. No
\emph{quantization} algorithm has been employed; the development of
the  quantum mechanical formalism has been autonomous, as promised.

From this point on, the development of QM along the traditional
lines can proceed.

For nonrelativistic particles with $m > 0$ and spin $s \geq 0$, we
have $\sch = L^2(\mathbb{R}^3, \mathbb{C}^{2s+1})$ and $\mathcal{D}
= \scs(\mathbb{R}^3, \mathbb{C}^{2s+1})$. The treatment of spin
being standard, we skip the details.

\vspace{.1in} \noindent  \emph{Remarks} (i) Note that the general
argument gives, in Eq.(30),  V(\textbf{X}, \textbf{P}) and not
V(\textbf{X}). In the next section we shall see that, for a quantum
system with the Hamiltonian (30) with V as a function of \textbf{X}
only, the classical Hamiltonian is the standard one given by
Eq.(94). It follows that, for systems for which the classical
situation is well described by a potential V(x), it is reasonable to
take, in the quantum Hamiltonian, the potential V(\textbf{X}).

\vspace{.1in} \noindent (ii) For particle motion in lower
dimensions, some of the fundamental observables are suppressed and
the system algebra is an appropriate subalgebra of the usual system
algebra ( say, $\sca^{(1)}$) for a particle moving in three
dimensions. For example, for a simple harmonic oscillator,  the
fundamental observables are $ X (= X_1)$ and $P (= P_1)$ (the
observables $X_2, X_3, P_2, P_3$ are suppressed); they, together
with the unit element I, generate a subalgebra $\sca_{osc}$ of
$\sca^{(1)}$. To identify the corresponding quantum triple
$(\hat{\sch}_{osc}, \hat{\mathcal{D}}_{osc}, \hat{\sca}_{osc})$, we
note that $\hat{\mathcal{D}}_{osc} = C^{\infty}(X,P) = \scs
(\mathbb{R})$ and $\hat{\sch}_{osc}$ is its completion
$L^2(\mathbb{R})$; $\hat{\sca}_{osc}$ is the algebra representing
$\sca_{osc}$ in the Schr$\ddot{o}$dinger representation. From the
remark (i) above, we have $ H = P^2/(2m) + (1/2) k X^2$  for the
quantum oscillator with X, P the traditional operators in the
Schr$\ddot{o}$dinger representation.

\vspace{.15in} \noindent \textbf{4. QUANTUM-CLASSICAL
CORRESPONDENCE}

\vspace{.12in} \noindent It will now be shown that supmech permits a
transparent treatment of quantum-classical correspondence. In
contrast to the general practice in this domain, we shall be careful
about the domains of operators and avoid some usual pitfalls in the
treatment of the $ \hbar \rightarrow 0 $ limit.

Our strategy will be to start with a quantum Hamiltonian system,
transform it to an isomorphic supmech Hamiltonian system involving
phase space functions and $ \star $-products [Weyl-Wigner-Moyal
formalism (Weyl [45], Wigner [47], Moyal [36])] and show that, in
this latter Hamiltonian system, the subclass of phase space
functions in the system algebra which go over to smooth functions in
the $\hbar \rightarrow 0$ limit yield the corresponding classical
Hamiltonian system. For simplicity, we restrict ourselves to the
case of a spinless nonrelativistic particle though the results
obtained admit trivial generalization to systems with phase space $
\mathbb{R}^{2n}$.

 In the existing literature, the works on quantum-classical
correspondence  closest to the present treatment are those of Liu
[32,33], Gracia-Bond$\acute{i}$a and V$\acute{a}$rilly [24] and
H$\ddot{o}$rmander [27]; some results from these works, especially
Liu [32,33], are used below [mainly in obtaining equations (93) and
(96)]. The reference (Bellissard and Vitot [7]) is a comprehensive
work reporting on some detailed features of quantum-classical
correspondence employing some techniques of noncommutative geometry;
its theme, however, is very different from ours.

In the case at hand, we have the quantum triple $(\sch, \mathcal{D},
\sca)$ where $ \sch = L^2(\mathbb{R}^3), \mathcal{D} =
\scs(\mathbb{R}^3)$ and \sca \ is the system algebra of a spinless
Galilean particle treated in section 3.4 as a standard quantum
system. As in Eq.(87), we shall take the potential function V to be
a function of $\mathbf{X}$ only. For $ A \in \sca $ and $ \phi,\psi
$ normalized elements in $\mathcal{D}$, we have the well defined
quantity
\begin{eqnarray*}
(\phi,A\psi) = \int \int \phi^*(y) K_{A}(y,y^\prime) \psi(y^\prime)
dydy^\prime
\end{eqnarray*}
where the kernel $ K_A$ is a (tempered) distribution. Recalling the
definition of Wigner function [47,49] corresponding to the wave
function $ \psi $ :
\begin{eqnarray}
W_\psi(x,p) = \int_{\mathbb{R}^3} exp[-ip.y/\hbar] \psi(x +
\frac{y}{2}) \psi^*(x-\frac{y}{2})dy
\end{eqnarray}
and defining the quantity  $A_W(x,p)$ by
\begin{eqnarray}
A_W(x,p) = \int exp[-ip.y/\hbar]K_A(x + \frac{y}{2}, x -
\frac{y}{2})dy
\end{eqnarray}
(note that $W_\psi$ is nothing but the quantity $P_W$ where P is the
projection operator $|\psi><\psi|$ corresponding to  $\psi$) we have
\begin{eqnarray}
(\psi,A\psi) = \int \int A_W(x,p) W_\psi(x,p) dx dp.
\end{eqnarray}
Whereas the kernels $K_A$ are distributions, the objects $A_W$ are
well defined functions. For example,
\begin{eqnarray*}
A & = & I : \hspace{.15in} K_A(y,y^\prime) = \delta(y-y^\prime)
\hspace{.15in}
A_W(x,p) = 1 \\
A & = & X_j  : \hspace{.15in} K_A(y,y^\prime) =
y_j\delta(y-y^\prime)
\hspace{.15in} A_W(x,p) = x_j \\
A & = & P_j : \hspace{.15in} K_A(y,y^\prime) =
-i\hbar\frac{\partial} {\partial y_j}\delta(y-y^\prime)
\hspace{.15in} A_W(x,p) = p_j.
 \end{eqnarray*}

The Wigner functions $ W_{\psi}$ are generally well-behaved
functions. We shall use Eq.(90) to characterize the class of
functions $A_W$ and call them Wigner-Schwartz integrable (WSI)
functions [i.e. functions integrable with respect to the Wigner
functions corresponding to the Schwartz functions in the sense of
Eq.(88)]. For the relation of this class to an appropriate class of
symbols in the theory of pseudodifferential operators, we refer to
Wong [49] and references therein.

The operator A can be reconstructed (as an element of \sca) from the
function $A_W$; for arbitrary $\phi, \psi \in \mathcal{D}$, we have
\begin{equation}
\begin{array}{l}
(\phi,A\psi) = \\
\displaystyle (2\pi\hbar)^{-3}\int \int \int exp[ip.(x-y)/\hbar]
\phi^*(x)A_W(\frac{x+y}{2},p)\psi(y)dpdxdy.
\end{array}
\end{equation}

 Replacing, on the right hand side of Eq.(88), the quantity $\psi(x
 +\frac{y}{2})\psi^*(x-\frac{y}{2})$ by $K_{\rho}(x+\frac{y}{2},
 x-\frac{y}{2})$ where $K_{\rho}(.,.)$ is the kernel of the density
 operator $\rho$, we obtain the Wigner function $\rho_W(x,p)$
 corresponding to $\rho$. Eq.(90) then goes over to the more
 general equation
 \begin{eqnarray} Tr(A \rho)  = \int \int A_W(x,p) \rho_W(x,p)dxdp.
 \end{eqnarray} The Wigner function $\rho_W$ is real but generally
 not non-negative.

  Introducing, in $ \mathbb{R}^{6}, $ the notations $ \xi $ = (x,p), $ d\xi = dxdp $
and $ \sigma ( \xi, \xi^{'} )
 = p.x^{'} - x.p^{'} $ (the symplectic form in $ \mathbb{R}^{6} $ ), we have, for
 A,B $ \in \sca $

\vspace{2mm} \noindent
\begin{eqnarray}
(AB)_{W}(\xi) & =  &     (2\pi)^{-6} \int \int  exp [ -i \sigma
                               (\xi - \eta, \tau)] A_{W} ( \eta +
                               \frac{\hbar \tau}{4} ). \nonumber \\
& \ & . B_{W} (\eta - \frac{\hbar \tau }{4})d\eta d\tau \nonumber \\
                        &  \equiv &  (A_{W} \star B_{W}) ( \xi).
\end{eqnarray}

The product $ \star $ of Eq.(93) is the \emph{twisted product} of
Liu [32,33] and the \emph{$\star$- product} of Bayen et al [6]. The
associativity condition $ A(BC) = (AB)C $ implies the corresponding
condition $ A_W \star (B_W \star C_W) = (A_W \star B_W) \star C_W $
in the space $ \sca_W$ of WSI functions which is a complex
associative non-commutative, unital *-algebra (with the star-product
as product and complex conjugation as involution). There is an
isomorphism between the two star-algebras \sca \ and $\mathcal{A}_W$
as can be verified from equations (93) and (91).

Recalling that, in the quantum Hamiltonian system $(\sca,
\omega_Q,H)$ the form $\omega_Q$ is fixed by the algebraic structure
of \sca \ and noting that, for the Hamiltonian H of Eq.(30) [with V
= V(\textbf{X})],
\begin{eqnarray}
H_W(x,p) = \frac{p^2}{2m} + V(x),
\end{eqnarray}
we have an isomorphism between the supmech Hamiltonian systems
$(\sca,\omega_Q,H) $ and $ (\sca_W,  \omega_W, H_W)$ where $\omega_W
= -i\hbar \omega_c^{(W)};$ here  $\omega_c^{(W)}$ is the canonical
2-form of the algebra $\sca_W$. Under this isomorphism, the quantum
mechanical PB (36) is mapped to the Moyal bracket
\begin{eqnarray}
\{ A_{W}, B_{W} \}_{M} \equiv (-i\hbar)^{-1} ( A_{W} \star B_{W} -
B_{W} \star A_{W} ).
\end{eqnarray}

  For  functions f, g in $\mathcal{A}_{W} $ which are smooth and
such that   $ f(\xi)$ and $ g(\xi)$ have no $ \hbar-$dependence, we
have, from Eq.(93),
\begin{eqnarray}
f \star g = fg - (i\hbar/2) \{ f, g \}_{cl} + O ( \hbar^{2} ).
\end{eqnarray}
The functions $ A_{W} (\xi) $ will have, in general, some $ \hbar $
dependence and the $ \hbar \rightarrow 0 $ limit may be singular for
some of them (Berry [9]). We denote by $(\mathcal{A}_{W})_{reg}$ the
subclass of functions in $ \mathcal{A}_{W} $ whose $ \hbar
\rightarrow 0 $ limits exist and are smooth (i.e. $ C^{\infty} $ )
functions; moreover, we demand that the Moyal bracket of every pair
of functions in this subclass also have smooth limits. This class is
easily seen to be a subalgebra of $ \mathcal{A}_{W}$ closed under
Moyal brackets.  Now, given two functions $A_W$ and $B_W$ in this
class, if $ A_{W} \rightarrow A_{cl} $
 and $ B_{W} \rightarrow B_{cl} $  as $ \hbar \rightarrow 0 $ then
 $ A_{W}  \star B_{W}
\rightarrow A_{cl} B_{cl} $; the subalgebra
$(\mathcal{A}_{W})_{reg}$, therefore, goes over, in the $ \hbar
\rightarrow 0 $ limit , to a subalgebra $\mathcal{A}_{cl}$ of    the
commutative algebra $ C^{\infty}(\mathbb{R}^{6}) $ (with pointwise
product as multiplication). The Moyal bracket of Eq.(95) goes over
to the classical PB $\{ A_{cl}, B_{cl} \}_{cl}$; the subalgebra
$\mathcal{A}_{cl}$, therefore, is closed under the classical Poisson
brackets. The classical PB $\{, \}_{cl}$ determines the
nondegenerate classical symplectic form $\omega_{cl}$. [ If $ \{ f,
g \}_{cl} = \sigma^{\alpha \beta} \frac{\partial f}{\partial
\xi^{\alpha}} \frac{\partial g}{\partial \xi^{\beta}}$, then
$\omega_{cl} = \sigma_{\alpha \beta} d \xi^{\alpha} \wedge d
\xi^{\beta}$ where the matrix $( \sigma_{\alpha \beta}) $ is the
inverse of the matrix $(\sigma^{\alpha \beta})$.] When $ H_{W} \in
(\mathcal{A}_{W})_{reg}$ [which is the case for the $H_W$ of
Eq.(94)],  the subsystem $ (\mathcal{A}_{W}, \omega_{W}, H_{W}
)_{reg}$ goes over to the supmech Hamiltonian system $
(\mathcal{A}_{cl}, \omega_{cl}, H_{cl})$.

When the $\hbar \rightarrow 0$ limits of $A_W$ and $\rho_W$ on the
right hand side of Eq.(92) exist (call them $A_{cl}$ and
$\rho_{cl}$), we have
\begin{eqnarray} Tr(A \rho) \rightarrow \int \int A_{cl}(x,p)
\rho_{cl}(x,p) dxdp. \end{eqnarray} The quantity $\rho_{cl}$ must be
non-negative (and, therefore, a genuine density function). To see
this, note that, for any operator $A \in \sca$ such that $A_W \in
(\sca_W)_{reg}$, the object $A^*A$ goes over to $\bar{A}_W* A_W$ in
the Weyl-Wigner-Moyal formalism which, in turn, goes to
$\bar{A}_{cl}A_{cl}$ in the $\hbar \rightarrow 0$ limit; this limit,
therefore, maps non-negative operators to non-negative functions.
Now if, in Eq.(97), A is a non-negative operator, the left hand side
is non-negative for an arbitrarily small value of $\hbar$ and,
therefore, the limiting value on the right hand side must also be
non-negative. This will prove the non-negativity of $\rho_{cl}$ if
the objects $A_{cl}$ in Eq.(97) realizable as classical limits
constitute a dense set of non-negative functions in $C^{\infty}(M)$.
This class is easily seen to include non-negative polynomials; good
enough.

In situations where the $\hbar \rightarrow 0$ limit of the time
derivative equals the time derivative of the classical limit [i.e.
we have $A(t) \rightarrow A_{cl}(t)$ and $\frac{dA(t)}{dt}
\rightarrow \frac{d A_{cl}(t)}{dt}$], the Heisenberg equation of
motion for A(t) goes over to the classical Hamilton's equation for
$A_{cl}(t)$. With a similar proviso, one obtains the classical
Liouville equation for $\rho_{cl}$ as the classical limit of the von
Neumann equation.

\vspace{.1in} Before closing this section, we briefly discuss an
interesting point :

  For commutative algebras, the inner derivations vanish and one can
have only outer derivations. Classical mechanics employs a subclass
of such algebras (those of smooth functions on  manifolds). It is an
interesting contrast to note that, while the quantum symplectics
employ  only inner derivations, classical symplectics employ only
outer derivations. The deeper significance of this is related to the
fact that the noncommutativity of  quantum algebras is generally
tied to the nonvanishing of the Planck constant $\hbar$. [This is
seen most transparently in the star  product of Eq.(93) above.] In
the limit $\hbar \rightarrow 0$, the algebra becomes commutative
(the star product of functions reduces to ordinary product) and the
inner derivations become outer derivations (commutators go over to
classical Poisson brackets implying that an inner derivation $D_A$
goes over to the Hamiltonian vector field $X_{A_{cl}}$).

\vspace{.15in} \noindent \textbf{5. AXIOMS}

\vspace{.12in} We shall now write down a set of axioms covering the
work presented in papers I and II. Before  the statement of axioms,
a few points are in order :

\vspace{.1in} \noindent (i) These axioms are meant to be
provisional; the `final' axioms will, hopefully, be formulated (not
necessarily by the present author) after a reasonably satisfactory
treatment of quantum theory of fields and space-time geometry in an
appropriately augmented supmech type framework  has been given.

\noindent (ii) The terms `system', `observation', `experiment' and a
few other `commonly used' terms will be assumed to be understood.
The term `relativity scheme' employed below will be understood to
have its meaning as explained in section 2.5.

\noindent (iii) The `universe' will be understood as the largest
possible observable system containing every other observable system
as a subsystem.

\noindent (iv) By an \emph{experimentally accessible system} we
shall mean one whose `identical' (for all practical purposes) copies
are reasonably freely available for repeated trials of an
experiment. Note that the universe and its `large' subsystems are
not included in this class.

\noindent (v) The term `system' will, henceforth will normally mean
an experimentally accessible one. Whenever it is intended to cover
the universe and/or its large subsystems (this will be the case in
the first three axioms only), the term system$^*$ will be used.

\vspace{.1in} The axioms will be labeled as \textbf{A1,..., A7}.

\vspace{.1in} \noindent \textbf{A1.}(\emph{Probabilistic framework;
System algebra and
states}) \\
(a) \emph{System algebra; Observables}. A system$^*$ S has
associated with it a topological superalgebra $\sca = \sca^{(S)}$
satisfying the conditions stated in section 3.4 of I. (Its elements
will be denoted as A,B,...). Observables of S are elements of the
subset \oa \ of even Hermitian elements of \sca.

\noindent (b) \emph{States}. States of the system$^*$, also referred
to as the states of the system algebra \sca \ (denoted by the
letters $\phi, \phi^{\prime},..$), are defined as continuous
positive linear functionals on \sca \ which are normalized [i.e.
$\phi(I) =1$ where I is the unit element of \sca]. The set of states
of \sca \ will be denoted as \scs(\sca) and the subset of pure
states by $\sone (\sca)$. For any $A \in \mathcal{O}(\sca)$ and $
\phi \in \scs (\sca)$, the quantity $\phi(A)$ is to be interpreted
as the expectation value of A when the system is in the state
$\phi$.

\noindent (c) Expectation values of odd elements of \sca \ vanish in
every pure state (hence in every state).

\noindent (d) \emph{Compatible completeness of observables and pure
states}. The pair \linebreak $(\oa,\  \sone(\sca))$ satisfies the CC
condition described in section 2.2.

\noindent (e) \emph{Experimental situations and probabilities}. An
experimental situation (relating to observations on the system$^*$
S) has associated with it a positive observable-valued measure
(PObVM) as defined in section 2.1; it associates, with measurable
subset of a measurable space (the `value space' of for the
quantities being measured), objects called supmech events which have
measure-like properties. Given the system prepared in a state
$\phi$, the probability of realization of a supmech event $\nu (E)$
is $\phi(\nu (E))$. It is stipulated that all probabilities in the
formalism relating to outcomes in experiments must be of this type.

 \vspace{.12in} \noindent \textbf{A2.} \emph{Differential calculus;
 Symplectic structure}. The system algebra \sca \ of a system$^*$ S is such as
to permit the development of superderivation-based differential
calculus on it (as described in section 2 of I); moreover, it is
equipped with a real symplectic form $\omega$ thus constituting a
symplectic superalgebra $ (\sca, \omega)$ [more generally, a
generalized symplectic superalgebra $(\sca, \scx, \omega)$ when the
derivations are restricted to a distinguished Lie sub-superalgera
\scx \ of  the Lie superalgebra \sdera of the superderivations of
\sca].

\vspace{.12in} \noindent \textbf{A3.} \emph{Dynamics}. The dynamics
of a system$^*$ S is described by an equicontinuous one-parameter
family of canonical transformations [satisfying the $C_0$ condition
(I, section 2.3)] generated by an even Hermitian
element H (the Hamiltonian) of \sca \ which is bounded below in the
sense that its expectation values in all pure states (hence in all
states) are bounded below.

\vspace{.12in} The mechanics described by the above-stated axioms
will be referred to  as Supmech. The triple $(\sca, \omega, H)$ or,
more precisely,  the quadruple $(\sca, \sone(\sca),\omega, H)$ will
be said to constitute a supmech Hamiltonian system.

\vspace{.12in} \noindent \textbf{A4.} \emph{Relativity scheme}. For
systems admitting space-time description, the `principle of
relativity', as described in section 2.5, will be operative.

\vspace{.12in} \noindent \textbf{A5.} \emph{Elementary systems;
Material particles}. (a) In every relativity scheme, material
particles will  be understood to be localizable elementary systems
(as defined in sections 2.4 and 2.5).

\noindent (b) The system algebra for a material particle will be the
one generated by its fundamental observables (as defined in section
2.5) and the identity element.

\vspace{.12in} \noindent \textbf{A6.} \emph{Coupled systems}. Given
two systems $S_1$ and $S_2$ described as supmech Hamiltonian systems
$(\sca^{(i)}, \scs_1^{(i)}, \omega^{(i)}, H^{(i)})$ (i=1,2), the
coupled system $(S_1 + S_2)$ will be described as a supmech
Hamiltonian system $(\sca, \scs_1, \omega, H)$ with
\[ \sca = \sca^{(1)} \hat{\otimes} \sca^{(2)}, \ \ \scs_1 = \scs_1(\sca),
 \ \ \omega = \omega^{(1)} \otimes I_2  + I_1 \otimes \omega^{(2)}
\] [where the symbol $\hat{\otimes}$ denotes the completed (skew)
tensor product and $I_1$ and $I_2$ are the unit elements of $\sca^{(1)}$
and  $\sca^{(2)}$ respectively] and H as in Eq.(100) of I.

\vspace{.1in} \noindent \emph{Note.} Theorem (2) in I implied
restrictions on the possible situations when the interaction of two
systems along the lines of the axiom \textbf{A6} can be consistently
described. A consequence of this theorem is that \emph{all}
experimentally accessible systems in nature must have either
supercommutative or non-supercommutative system algebras. The next
axiom indicates the choice.

\vspace{.12in} \noindent \textbf{A7.} \emph{Quantum systems}.  All
(experimentally accessible) systems in nature have
non-supercommutative system algebras (and hence are quantum
systems); they have a quantum symplectic structure (as defined in
section 3.3 of I) with the universal parameter $b = -i \hbar$.

\vspace{.1in} \noindent \emph{Note.} (i) The quantum systems  were
shown (in section 3.2) to have equivalent (as supmech Hamiltonian
systems) Hilbert space based realizations (without introducing
additional postulates); those having finitely generated system
algebras were guaranteed to have their system algebras represented
irreducibly in the Hilbert space.

\vspace{.1in} \noindent (ii) Axioms A7 and A5(a) imply that all
material particles are localizable elementary quantum systems. Since
they have finitely generated system algebras, the corresponding
supmech Hamiltonian systems are guaranteed to have Hilbert space
based realizations with the system algebra represented faithfully
and irreducibly. They can be treated as in section 3.4 without
introducing any extra postulates; in particular, introduction of the
Schr$\ddot{o}$dinger wave functions with the traditional Born
interpretation and the Schr$\ddot{o}$dinger dynamics are automatic.

\vspace{.1in} \noindent (iii) General quantum systems were shown in
section 3.2 to admit commutative superselection rules.

 \vspace{.15in} \noindent \textbf{6. CONCLUDING REMARKS}

\vspace{.12in} \noindent 1. The central message of the first two
papers in this series is this : Complex associative algebras are the
appropriate objects for the development of a universal mechanics.
The proposed universal mechanics--- supmech --- is constrained by
the formalism (and empirical acceptability) to reduce to traditional
quantum mechanics for all `experimentally accessible' systems. It is
worth re-emphasizing that, for an autonomous development of quantum
mechanics, the fundamental objects are algebras and not Hilbert
spaces.

\vspace{.12in} \noindent 2. A contribution of the present work
expected to be of some significance for the algebraic schemes in
theoretical physics and probability theory is  the introduction of
the condition of compatible completeness for observables and pure
states [the CC condition : axiom A1(d)] which plays an important role
in ensuring that
the  quantum systems defined algebraically in section 3.1, have
faithful Hilbert space-based realizations. It is desirable to
formulate necessary and/or sufficient conditions on the superalgebra
\sca \ alone (i.e. without reference to states) so that the CC
condition is automatically satisfied.

An interesting result, obtained in section 2.3, is that the
superclassical systems with a finite number of fermionic generators
generally do not satisfy the CC condition. This probably explains
their non-occurrence in nature. It is worth investigating whether
the CC condition is related to some stability property of dynamics.

\vspace{.12in} \noindent 3. Some features of the development of QM
in the present work (apart from the fact that it is autonomous)
should please theoreticians : there is a fairly broad-based
algebraic formalism connected smoothly to the Hilbert space QM;
there is a natural place for commutative superselection rules and
for the Dirac's bra-ket formalism; the Planck constant is introduced
`by hand' at only one place (at just the right place : the quantum
symplectic form) and it appears at all conventional places
automatically. Moreover, once the concepts of localization,
elementary system and standard quantum system  are introduced at
appropriate places, it is adequate to define a material particle as
a localizable elementary quantum system ; `everything else' ---
including the emergence  of the Schr$\ddot{o}$dinger wave functions
with their traditional interpretation and the Schr$\ddot{o}$dinger
equation --- is automatic.

\vspace{.12in} \noindent 4. The treatment of quantum-classical
correspondence in section 4, illustrated with the example of a
nonrelativistic spinless particle, makes clear as to how the subject
should be treated in the general case :  go from the traditional
Hilbert space -based description of the quantum system to an
equivalent (in the sense of a supmech hamiltonian system) phase
space description in the Weyl-Wigner-Moyal formalism, pick up the
appropriate subsets in the observables and states having smooth
$\hbar \rightarrow 0$ limits and verify that the limit gives a
commutative  supmech Hamiltonian system (which is generally a
traditional classical hamiltonian system).

\vspace{.2in} \noindent \textbf{ACKNOWLEDGEMENTS}

The author thanks K.R. Parthasarathy for helpful discussions.

\vspace{.2in} \noindent \textbf{REFERENCES} {\footnotesize
\begin{description}
\item[[1]] V. Aldaya, J.A. De Azcarraga, Geometric formulation of
classical mechanics and field theory, Nuov. Cim. \textbf{3} (1980)
1-66.
\item[[2]] L.M. Alonso,  Group Theoretical Foundations of Classical and
Quantum Mechanics. II. Elementary Systems, J. Math. Phys.
\textbf{20} (1979) 219-230.
\item[[3]] A.P. Antoine, J. Math. Phys. \textbf{10} (1969) 53-69,2276-2290.
\item[[4]] H. Bacry,  Localizability and Space in Quantum Physics,
Lecture Notes in Physics, vol 308, Springer-Verlag, Berlin, 1988.
\item[[5]] V. Bargmann, On unitary ray representations of
continuous groups, Ann. Math. \textbf{59} (1954) 1-46.
\item[[6]] F. Bayen et al, Deformation theory and quantization
(I. Deformations of symplectic structures; II. Physical
applications), Annals of Phys. \textbf{110} (1978) 61,111.
\item[[7]] J. Bellissard, M. Vitot, Heisenberg's picture and
non-commutative geometry of the semiclassical limit in quantum
mechanics, Ann. Inst. Henri Poincar$\acute{e}$ \textbf{52} (1990)
175.
\item[[8]] F.A. Berezin, Superanalysis, edited by A.A. Kirillov,
 D. Reidel Pub. Co., Dordrecht, 1987.
\item[[9]] M. Berry,  `Some Quantum-Classical Asymptotics' in
Chaos and Quantum Physics,  Les Houches, session LII, 1989, J.
Elsevier Science Publishers, 1991.
\item[[10]] N.N. Bogolubov, A.A. Logunov, I.T. Todorov,
Introduction to Axiomatic  Quantum Field Theory, Benjamin/Cummings,
Reading, 1975.
\item[[11]] A. B$\ddot{o}$hm, The Rigged Hilbert Space and Quantum
Mechanics, Lecture Notes in Physics, vol 78, Springer, Berlin, 1978.
\item[[12]] P. Busch, M. Grabowski, P.J. Lahti, Operational Quantum
Physics, Springer-Verlag, Berlin, 1995.
\item[[13]] J.F. Cari$\tilde{n}$ena, M. Santander, On the Projective Unitary
Representations of Connected Lie Groups, J. Math. Phys. \textbf{16}
(1975) 1416-1420.
\item[[14]] T. Dass,  Symmetries, gauge fields, strings and
fundamental interactions, vol. I: Mathematical techniques in gauge
and string theories, Wiley Eastern Limited, New Delhi, 1993.
\item[[15]] T. Dass,  A Stepwise Planned Approach to the Solution of
Hilbert's Sixth Problem. I : Noncommutative Symplectic Geometry and
Hamiltonian Mechanics. arXiv : 0909.4606  [math-ph] (2009).
\item[[16]] T. Dass, S.K. Sharma, Mathematical Methods in
Classical and Quantum Physics. Universities Press, Hyderabad, 1998.
\item[[17]] E.B. Davies, Quantum Theory of Open
Systems, Academic Press, London, 1976.
\item[[18]] R. de la Madrid, The role of the rigged Hilbert space in
quantum mechanics. \emph{Eur. J. Phys.} \textbf{26} (2005) 287-312;
ArXiv : quant-ph/0502053.
\item[[19]] C. DeWitt-Morette, K.D. Elworthy,  A stepping stone to
stochastic analysis, Phys. Rep. \textbf{77} (1981) 125-167.
\item[[20]] D.A. Dubin, M.A. Hennings Quantum Mechanics, Algebras
and Distributions,  Longman Scientific and Technical, Harlow, 1990.
\item[[21]] M. Dubois-Violette,  `Lectures on Graded Differential Algebras and
Noncommutative Geometry' in Noncommutative Differential Geometry and
its Aapplication to Physics (Shonan, Japan, 1999), pp 245-306.
Kluwer Academic Publishers, 2001; arXiv: math.QA/9912017.
\item[[22]] I.M. Gelfand, N.J. Vilenkin, Generalized Functions,
vol. IV, Academic Press, New York, 1964.
\item[[23]]  J. Glimm, A. Jaffe, Quantum Physics: a Functional
Integral Point of View, Springer Verlag, New York, 1981.
\item[[24]] J.M. Gracia-Bond$\acute{i}$a, J.C. V$\acute{a}$rilly,
Phase space representation for Galilean quantum particles of
arbitrary spin, J.Phys.A: Math.Gen. \textbf{21} (1988) L879-L883.
\item[[25]] V. Guillemin, S. Sternberg, Symplectic Techniques in
Physics, Cambridge University Press, 1984.
\item[[26]] A.S. Holevo, Probabilistic and Statistical Aspects of
Quantum Theory, North Holland Publishing Corporation, Amsterdam,
1982.
\item[[27]] L. H$\ddot{o}$rmander, Weyl calculus of pseudodifferential
operators, Comm. Pure Appl. Math. \textbf{32} (1979) 359-443.
\item[[28]] S.S. Horuzhy, Introduction to Algebraic Quantum Field
Theory, Kluwer Academic Publishers, Dordrecht, 1990.
\item[[29]] A. Inoue, Tomita-Takesaki Theory in Algebras of Unbounded
Operators, Springer, Berlin, 1998.
\item[[30]] P. Kristensen, L. Mejlbo, E. Thue Poulsen, Tempered
distributions in infinitely many dimensions I. Canonical field
operators, Comm. Math. Phys. \textbf{1} (1965) 175-214.
\item[[31]] G. Lassner, Algebras of unbounded operators and quantum
dynamics, Physica \textbf{124A} (1984) 471-480.
\item[[32]] K.C. Liu, J. Math. Phys.  \textbf{16} (1975) 2054.
\item[[33]] K.C. Liu, J. Math. Phys. \textbf{17} (1976) 859.
\item[[34]] G.W. Mackey, Imprimitivity for representations of
locally compact groups, Proc. Nat. Acad. Sci. U.S. \textbf{35}
(1949) 537-545.
\item[[35]] Y. Matsushima, Differentiable Manifolds, Marcel Dekker,
New York, 1972.
\item[[36]] J.E. Moyal, Quantum mechanics as a statistical theory,
Proc. Camb. Phil. Soc. \textbf{45} (1949) 99-124.
\item[[37]] T.D. Newton, E.P. Wigner, Localized states for
elementary systems, Rev. Mod. Phys. \textbf{21} (1949) 400-406.
\item[[38]] K.R. Parthasarathy, An Introduction to Quantum
Stochastic Calculus, Birkha$\ddot{u}$ser, Basel, 1992.
\item[[39]] R.T. Powers, Self-adjoint algebras of unbounded
operators, \emph{Comm. Math. Phys.} \textbf{21} (1971) 85-124.
\item[[40]] J.E. Roberts, J. Math. Phys.
\textbf{7} (1966) 1097-1104.
\item[[41]] W. Rudin, Functional Analysis,  Tata
McGraw-Hill, New Delhi, 1974.
\item[[42]] J.-M. Souriau, Structure of Dynamical Systems,
a Symplectic View of Physics,  Birkh$\ddot{a}$user, Boston, 1997.
\item[[43]] E.C.G. Sudarshan, N. Mukunda  \emph{Classical Dynamics : A
Modern Perspective}.  Wiley, New York, 1974.
\item[[44]] V.S. Varadarajan, Geometry of Quantum Theory, 2nd
ed.,  Springer-Verlag, New York, 1985.
\item[[45]] H. Weyl, Theory of Groups and Quantum
Mechanics, Dover, New York, 1949.
\item[[46]] A.S. Wightman, On the localizability of quantum
mechanical systems, Rev. Mod. Phys. \textbf{34} (1962) 845-872.
\item[[47]] E.P. Wigner, On the quantum correction for thermodynamic
equilibrium, Phys. Rev. \textbf{40} (1932) 749-759.
\item[[48]] E.P. Wigner, Unitary representations of the
inhomogeneous Lorentz group, Ann. Math.(N.Y.) \textbf{40} (1939)
149-204.
\item[[49]] M.W. Wong, Weyl Transforms, Springer, New York, 1998.

\end{description}}

\end{document}